\documentclass[preprint,prc,showpacs,showkeys,longbibliography]{revtex4-1}
\usepackage{epsfig,graphicx,float,color}
\usepackage{amssymb}
\usepackage{amsmath}
\usepackage[normalem]{ulem}

\begin{document}
\title{Shape coexistence in Sr isotopes}
\author{E.~Maya-Barbecho$^{1}$ and J.E.~Garc\'{\i}a-Ramos$^{1,2}$}
\affiliation{
$^1$ Departamento de  Ciencias Integradas y Centro de Estudios Avanzados en F\'isica, Matem\'atica y Computaci\'on, Universidad de Huelva,
  21071 Huelva, Spain\\
$^2$ Instituto Carlos I de F\'{\i}sica Te\'orica y Computacional,  Universidad de Granada, Fuentenueva s/n, 18071 Granada, Spain
} 
\begin{abstract} 
\begin{description}
\item [Background]   
  Sr isotopes are located in the mass region $A\approx 100$, where a very quick onset of nuclear deformation exists, being other notable examples of this area Yb, Zr, and Nb nuclei. The presence of the proton subshell closure $Z=40$ allows the existence of particle-hole excitations that produces low-lying intruder bands.
\item [Purpose]
  The goal of this work is the study of the nuclear structure of the even-even $^{92-102}$Sr isotopes through the accurate description of excitation energies, $B(E2)$ transition rates, nuclear radii and two-neutron separation energies.   
\item [Method] 
  The interacting boson model with configuration mixing will be the framework to calculate all the observables of the Sr isotopes. Only two types of configurations will be considered, namely, 0particle-0hole and 2particle-2hole excitations. The parameters of the model are determined using a least-squares procedure for the excitation energies and the $B(E2)$ transition rates.
\item [Results]
  For the whole chain of isotopes, the value of excitation energies, $B(E2)$'s, two-neutron separation energies, nuclear radii, and isotope shifts have been obtained, with a good agreement between theory and experiment. Also, a detailed analysis of the wave functions have been performed and, finally, the mean-field energy surfaces and the value of the nuclear deformation, $\beta$, have been obtained.   
\item [Conclusions]
  The presence of low-lying intruder states in even-even Sr isotopes have been confirmed and its connection with the onset of deformation has been clarified. Lightest Sr isotopes present a spherical structure while the heaviest ones are clearly deformed. The rapid onset of deformation at neutron number $60$ is due to the crossing of the regular and intruder configurations and, moreover, both families of states present an increase of deformation with the neutron number. 
\end{description}
\end{abstract}


\date{\today}
\maketitle

\section{Introduction}
\label{sec-intro}
Nuclear deformation plays a major role in the understanding of the evolution of nuclear structure along the mass table.  The shape of the nucleus is determined by a fine balance between, on one hand, the stabilizing effect of closed shells that tends to make the nucleus to exhibit properties of spherical shape, and, on the other, the pairing and the quadrupole force, that tends to make the nucleus quadrupole deformed \cite{Otsu18}. This balance generates regions near to closed shell with spherical shapes, while those around mid-shell are well deformed. In particular, the evolution of the single particle levels (monopole term) and the strength of the rest of low multipolar terms of the nuclear interaction are key elements for a correct understanding of the nuclear shape. The correct description of the transition from one limit  to the other remains at present a major issue in theoretical nuclear physics.

The situation is even more appealing when in the nuclear spectrum coexist states with different shapes in a  narrow energy range. This phenomenon is named shape coexistence and in nuclear physics it was first proposed by Morinaga \cite{morinaga56} to explain the nature of the first excited state in $^{16}$O, a $0^+$ state which was assumed to be deformed though the ground state is obviously spherical because the nucleus is doubly magic. This state corresponds to a proton 2p-2h excitation across the $Z=8$ shell closure and, therefore, it is a intruder state with a deformed structure. This idea was also supported by the early works \cite{Brown64,Brown66a,Brown66b} for oxigen isotopes. In closed nuclei, such as $^{16}$O, both proton and neutron excitations across magic number $8$ appear as the major component of the structure of the first $0^+$ state and lead naturally, through the effect of the quadrupole part of the nucleon-nucleon interactions, to the existence of a deformed band built on top of this $0^+$ (called intruder). This becomes a major indication for the appearance of shapes with an unexpected feature in nuclei that contain doubly magic proton and neutron configurations. One of the most explicit example of such shape coexistence is observed experimentally in $^{40}$Ca and is very well described through shell model calculations involving mp-nh excitations\cite{Caur07,Poves16,Poves18}.

Experimentally, the most dramatic example of shape coexistence is provided by the systematics of the nuclear radii and the isotope shifts observed in certain mass regions. The existence of sudden changes in that observables hint to the presence of intruder states that affect the deformation of the ground state. In particular, that was first observed in the case of Hg where the odd-even staggering in the radius systematics clearly points to the presence of states with rather different degree of deformation. Since then, shape coexistence has been found to play a major role to explain the nuclear structure in many mass regions, specially in those close to a shell or subshell closure in protons (neutrons) and around the mid-shell in neutrons (protons), being present in light, medium, and heavy nuclei \cite{hey83,wood92,heyde11}.

The theoretical description of shape coexistence can be done with two complementary approaches, namely, the nuclear shell model and the self-consistent way, based on Hartree-Fock (HF) or Hartree-Fock-Bogoliubov (HFB) theories. The residual interaction used in the shell-model Hamiltonian has been refined in the last decades as shown in \cite{Brow06} where a new universal interaction for the $sd$ shell has been designed or in  \cite{Nowa09} where the interaction has been obtained for the $sd-pf$ shell. The description of the region around $Z=40$ is driven by the simultaneous occupation of neutrons and protons in spin-orbit partners. Hence, once the neutron $1g_{7/2}$ orbit started to be filled, the interaction with the proton $1g_{9/2}$ orbit favours the existence of a zone of deformation in Zr and Sr nuclei with neutron number larger than $58$. This idea was proposed in the seminal works of Federman and Pittel \cite{Fede77,Fede79a,Fede79b}, entitled ``A unified shell-model description of nuclear deformation'' where the key importance of the simultaneous occupation of the neutron-proton spin-orbit partners is emphasized. Federman and coworkers explored in depth this mass area using a reduced model space consisting of the $3s_{1/2}$, $2d_{3/2}$, and $1g_{7/2}$ neutron orbits and, in the case of protons, the $2p_{1/2}$, $1g_{9/2}$, and $2d_{5/2}$ orbits \cite{Fede84,Heyd88,Etch89,Pitt93}. More recently, large scale shell-model calculations have been performed for the same mass region using more realistic valence spaces, as in the case of \cite{Holt00} or \cite{Siej09}. However, in order to correctly treat the possible existence of np-nh excitations across the $Z=40$ shell gap, it is needed to include other shells, therefore, rapidly increasing the size of the model space, running outside present computational capabilities. To overcome this limitation, the Monte Carlo Shell Model was introduced by \citet{Otsuka01,Shimizu12, Shimizu17} and it has been successfully applied to the Zr region \cite{Togashi16}. In this work, it was observed the filling of the proton $1g_{9/2}$ shell either for the ground state or for low-lying excited states. 

The other main-stream approach to deal with shape coexistence and configuration mixing is the mean-field approach. In \cite{Mei12}, the authors studied the rapid change in the structure of Sr an Zr isotopes solving a five-dimensional collective Hamiltonian with parameters coming from a non-relativistic Skyrme interaction and the PC-PK1 and SLy4 relativistic forces. A spherical-oblate-prolate shape transition in neutron-rich Sr and Zr isotopes is observed. In \cite{Xian12}, the authors studied Kr, Sr, Zr, and Mo isotopes around neutron number $60$ with the relativistic interaction PC-PK1. A rapid evolution in Sr and Zr while moderated in Mo and Kr is observed, in addition to a prolate-oblate shape coexistence in $^{98}$Sr and $^{100}$Zr. In \cite{Nomu16}, the even-even Ru, Mo, Zr, and Sr are studied within the HFB approach using a Gogny-D1M interaction. The different spectroscopic properties are obtained thanks to the mapping of the energy density functional into an interacting boson model with configuration mixing (IBM-CM) energy surface. In the case of Sr as in the case of Zr, there is a rapid change in the structure of the $0_1^+$ and $0_2^+$ states when approaching to neutron number  $60$. The spectroscopic properties of Zr and Sr isotopes, among many others, have been obtained in \cite{Dela10} using a generalized Bohr Hamiltonian which parameters are fixed through a reduction of the Hill-Wheeler generator coordinate method equation using a Gaussian overlap approximation (GOA). In \cite{Paul17}, the experimental information for Zr isotopes is compared with the theoretical results obtained with three different approaches, namely, the GOA reduction to the five-dimensional Bohr Hamiltonian, the solutions of the Hill-Weeler equations using the SLyMR00 version of the Skyrme force, or using a Gogny D1S interaction. All the approaches provide a good description of the energy systematics of the $2_1^+$ state, but differences are shown up for the energy ratio $E(4_1^+)/E(2_1^+)$. Calculations that represent the state-of-the-art of the field  were carried out by \citet{Rodr10} within the HFB framework, allowing to handle both the axial and triaxial degrees of freedom on equal footing, with full symmetry conserving configuration mixing calculations  and with application to Sr, Zr, and Mo.

The present work extends our previous analysis of the $Z\approx 40$ and $A\approx 100$ region \cite{Garc19, Garc20}, in particular, the even-even Zr isotopes, to the case of Sr. Sr is also an excellent candidate to study the influence of the intruder states on the onset of deformation for $N\approx 60$ because it presents a rapid lowering of the energy of the $2_1^+$ state, a rapid growing of the ratio $E(4_1^+)/E(2_1^+)$, a sudden increase in the radius or a flatness of the two-neutron separation energy.

The papers is organized as follows. In Section \ref{sec-exp}, the present experimental knowledge on Sr is reviewed; in Section \ref{sec-ibm-cm}, we present the theoretical framework of this work, i.e., the IBM-CM, including the procedure used to get the fitting parameters of the model; in Section \ref{sec-corr_energy}, the study of the correlation energy gain is presented; in Section \ref{sec-comp}, a detailed comparison of theory and experiment is presented; in Section \ref{sec-wf}, the analysis of the wave functions is given; in Section \ref{sec-other}, we study the radii, the isotopic shifts and the two-neutron separation energies; in Section \ref{sec-deformation}, we calculate the mean-field energy surfaces and deformations and, finally, in Section \ref{sec-conclu} the summary and the conclusions are presented. 

\section{Experimental data in the even-even Sr nuclei }
\label{sec-exp}
\begin{figure}[hbt]
\centering
\includegraphics[width=1.00\linewidth]{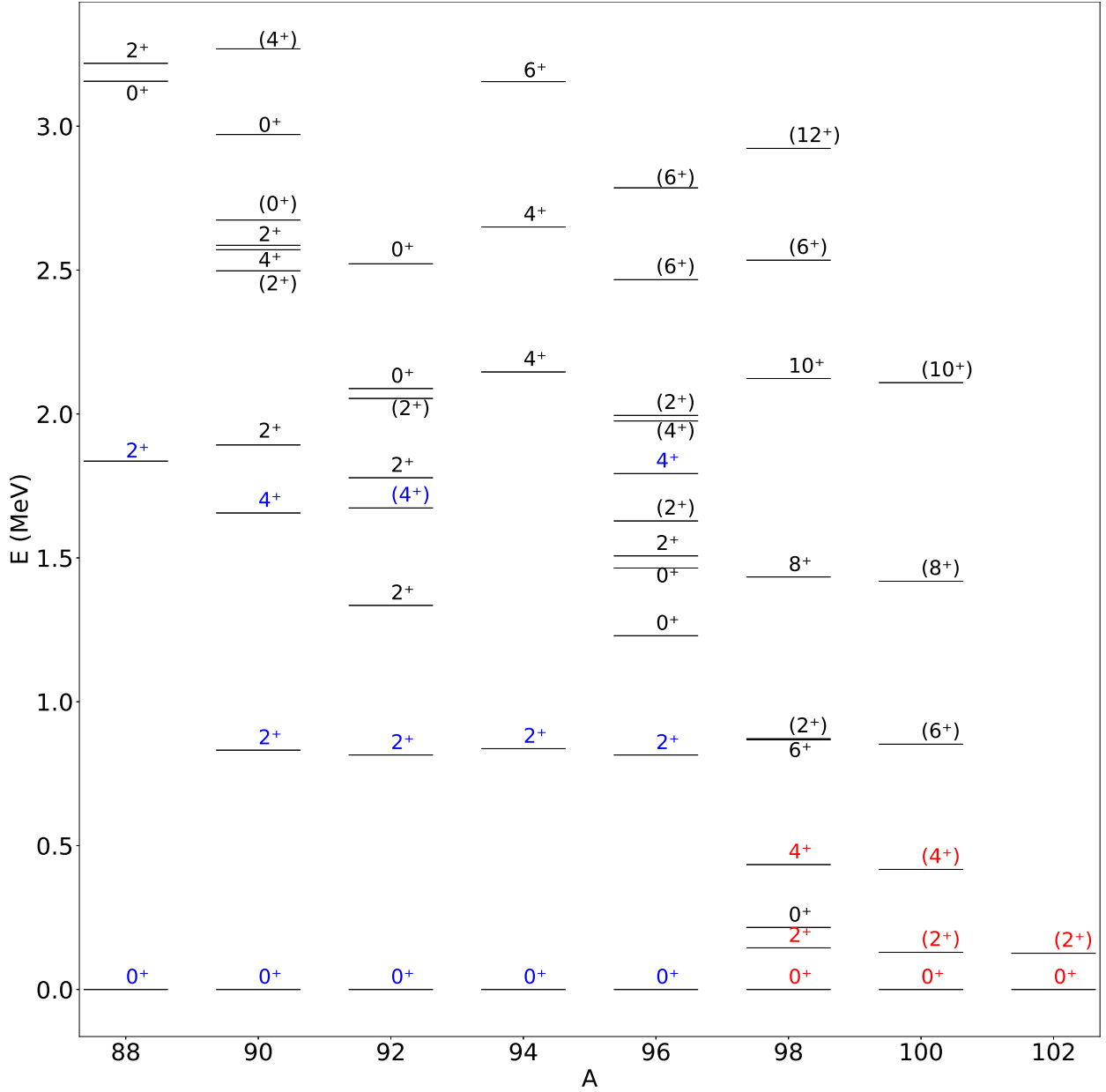}
\caption{Experimental energy level systematics of positive parity states for the Sr isotopes. Only levels up to $E_x \approx 3$ MeV are shown. Levels labelled in blue correspond to spherical shapes, while labelled in red to deformed ones (see information in Section \ref{sec-exp}).}
\label{fig-e-systematics-sr}
\end{figure}

The experimental knowledge for even-even Sr isotopes in the neutron $50-82$ shell spans from the closed-shell nucleus $^{88}$Sr till the deformed one $^{102}$Sr, as it is depicted in Fig.\ \ref{fig-e-systematics-sr}, where the energy systematics of positive-parity states up to an excitation energy $E\approx 3$ MeV is presented. These isotopes clearly need for more experimental data and, in particular, the use of complementary techniques to extract information on quadrupole moments using Coulomb excitation methods and lifetime measurements. The experimental data on $\rho^2 (E0)$ will lead to direct information 
concerning shape mixing as has been show in the nearby Zr isotopes
and many other regions.
In Fig.\ \ref{fig-e-systematics-sr} we used blue and red colors for the states which structure as been established firmly in the recent literature as being regular (vibrational like) or intruder (rotational like), respectively, and as it is discussed in detail below. The nucleus $^{88}$Sr is characterized by a large energy gap between the ground state and the first excited one, a $2^+$ state, as it corresponds to the shell closure; in $^{90-92}$Sr an underlying vibrational spectrum is seen although it becomes much more complex as the energy increases; in $^{94}$Sr the energy of the $2_1^+$ state remains constant but a large energy gap appears with the $4_1^+$ state; when passing from $^{96}$Sr to $^{98}$Sr the spectrum is suddenly compressed with a much smaller $E(2_1^+)$ and for  $^{100-102}$Sr only the yrast band is observed with a clear rotational structure.

For the comparison with the theoretical calculations, we have considered, when available, the evaluated experimental data appearing in Nuclear Data Sheets publications for $A=92$ \cite{Bagl12b}, $A=94$ \cite{Abri06,Negr11}, $A=96$ \cite{Abri08}, $A=98$ \cite{Sing03,Chen20}, $A=100$ \cite{Sing08}, and $A=102$ \cite{Defr09} complemented with the most up to date references for certain isotopes. Note that we do not consider in our analysis the nuclei $^{88-90}$Sr because they are at or too closed to the shell closure where our theoretical approach is not reliable enough. 

In \cite{Rzac09} the isotopes $^{92,94,96}$Sr have been studied through the spontaneous fission of $^{248}$Cm and new spin and parity were assigned to several energy level which differ from previous works. \citet{Park16} conducted a  gamma-ray and internal conversion electron spectroscopy experiment for $^{98}$Sr at the TRIUMF-ISAC facility. In this work, the $\rho^2(E0)$ value for the transition $0_2^+\rightarrow 0_1^+$ and the $B(E2)$ values for  $2_1^+\rightarrow 0_1^+$,  $0_2^+\rightarrow 2_1^+$, and  $4_1^+\rightarrow 2_1^+$ transitions were obtained and, except for the case of the $B(E2; 0_2^+\rightarrow 2_1^+$) transition, the obtained values were compatible with the evaluated ones. The obtained mixing for the first two $0^+$ states was $9\%$ while for the two $2^+$ states $1.3\%$. In \cite{Clem16}, the authors have investigated the neutron-rich $^{96,98}$Sr isotopes using Coulomb excitation at REX-ISOLDE facility (CERN). BE2's and spectroscopic quadrupole moments have been extracted from the cross sections. A small mixing is obtained between a spherical and a prolate configuration in $^{98}$Sr. 
In \cite{Clem17}, the authors conducted an experiment for neutron-rich Sr isotopes using Coulomb excitation with radiactive beams at REX-ISOLDE facility (CERN) using the MINIBALL spectrometer. They concluded that there is a sudden increase in the value of $\beta$ for $^{98}$Sr with a mixing between regular and intruder states of only a $12\%$. In \cite{Regi17,*Regi17erratum}, the authors measured the lifetime of the isotopes $^{94,96,98}$Sr using LaBr3(Ce) detectors combined with Ge ones in the EXILL-FATIMA spectrometer at the Institut Laue-Langevin in Grenoble. They obtained values for transition rates in the yrast band up to $J=6^+$, observing the rapid onset of deformation in $^{98}$Sr.
In \cite{Cruz18}, the spectroscopic factors of $0^+_{2,3}$ states were determined for $^{96}$Sr and the results are consistent with a $40\%$ mixing and a difference in $\beta$ between both $0^+$ states of $0.31$.
In \cite{Esma19}, delayed gamma rays in $A=97$ fission fragments were analyzed too at the high flux nuclear reactor of the Institute Laue-Langevin in Grenoble. Several lifetimes for $^{97}$Sr were measured using a fast-timing setup. This nucleus is at the spherical-deformed border, i.e., in between neutron number $58$ and $60$ that is the place where Sr isotopes pass from being spherical to deformed.
\citet{Urba19a} have studied the $0_2^+$ states of $^{98}$Sr and $^{100}$Zr analyzing the spontaneous fission of $^{248}$Cm and $^{252}$Cf in the Eurogam2 and the Gammasphere arrays, moreover, in this work they classified phenomenologically the $0^+$ states in four categories, concluding that the first two $0^+$ states are highly deformed and correspond to 2p-2h excitations of two neutrons from the $9/2^+ [404]$  extruder level to the low-$\Omega$ prolate and the $11/2^- [505]$ oblate orbitals originating from the $h_{11/2}$ shell, respectively.
More recently, the same group \cite{Urba21} reinvestigated the excited states of $^{90,92,94,96,98}$Sr using Exogam and Gammasphere arrays. $23$ new levels with $30$ new decays were obtained in four nuclei.
Finally, in \cite{Cruz20}, the authors studied the level structure of $^{93,94,95}$Sr via one-neutron stripping
reactions at TRIUMF. The results suggest a strong mixing in $^{94}$Sr and in $^{96}$Sr, showing indications of shape transition before $N=60$. It is of relevance the observations  in $^{94}$Sr of two $0^+$ states at $1880$ and $2293$ keV (not considered in this work)

\section{The Interacting Boson Model with configuration mixing formalism}
\label{sec-ibm-cm}
\subsection{The formalism}
\label{sec-formalism}
The IBM-CM supposes an extension of the original formulation of the IBM \cite{iach87} and it allows the simultaneous treatment of several boson configurations corresponding to particle-hole excitations across a shell or subshell closure \cite{duval81,duval82}. In this version of the model no distinction is made between proton and neutron bosons. In the case of Sr, this shell closure is assumed to exist for $Z=40$, corresponding the regular states to a 2h proton configuration, the intruder ones to a 4h-2p proton configuration and the number of valence neutrons is determined considering $50$ as the neutron closed shell. Therefore, the number of valence bosons, $N$, will be half of the sum of the valence protons, which is $2$, plus half the number of valence neutrons. The intruder configuration will own, in addition, $2$ extra bosons. Hence, the regular plus the intruder space will correspond to a $[N] \oplus [N+2]$ Hilbert space. The Hamiltonian will be made of two sectors, one corresponding to the regular part, $[N]$, another to the intruder one, $[N+2]$, and an interaction term, being written the total Hamiltonian as,
\begin{equation}
  \hat{H}=\hat{P}^{\dag}_{N}\hat{H}^N_{\rm ecqf}\hat{P}_{N}+
  \hat{P}^{\dag}_{N+2}\left(\hat{H}^{N+2}_{\rm ecqf}+
    \Delta^{N+2}\right)\hat{P}_{N+2}\
  +\hat{V}_{\rm mix}^{N,N+2}~,
\label{eq:ibmhamiltonian}
\end{equation}
where $\hat{P}_{N}$ and $\hat{P}_{N+2}$ are projection operators onto the $[N]$ and the $[N+2]$ boson spaces, respectively,
\begin{equation}
  \hat{H}^i_{\rm ecqf}=\varepsilon_i \hat{n}_d+\kappa'_i
  \hat{L}\cdot\hat{L}+
  \kappa_i
  \hat{Q}(\chi_i)\cdot\hat{Q}(\chi_i)
  \label{eq:cqfhamiltonian}
\end{equation}
is a simplified IBM Hamiltonian named extended consistent-Q Hamiltonian (ECQF), \cite{warner83,lipas85} with $i=N,N+2$, with $\hat{n}_d$ the $d$ boson number, $\hat{L}$ the angular momentum, and $\hat{Q}(\chi)$ the quadrupole operator. 
The parameter $\Delta^{N+2}$ represents the energy needed to excite two proton particles across the $Z=40$ shell gap, giving rise to 2p-2h excitations, corrected with the pairing interaction gain and including monopole effects \cite{Hey85,Hey87}. The operator $\hat{V}_{\rm mix}^{N,N+2}$ describes the mixing between the $N$ and the $N+2$ configurations and is defined as
\begin{equation}
  \hat{V}_{\rm mix}^{N,N+2}=\omega_0^{N,N+2}(s^\dag\times s^\dag + s\times
  s)+\omega_2^{N,N+2} (d^\dag\times d^\dag+\tilde{d}\times \tilde{d})^{(0)}.
\label{eq:vmix}
\end{equation}
In this work, we assume $\omega_0^{N,N+2}=\omega_2^{N,N+2}=\omega$.

The $E2$ transition operator is built with the same quadrupole operator that appears in the Hamiltonian (\ref{eq:cqfhamiltonian}) and it is defined as the sum of two contributions that act separately in the regular and the intruder sectors without crossed contributions, 
\begin{equation}
  \hat{T}(E2)_\mu=\sum_{i=N,N+2} e_i \hat{P}_i^\dag\hat{Q}_\mu(\chi_i)\hat{P}_i~.
  \label{eq:e2operator}
\end{equation}
The $e_i$ ($i=N,N+2$) are the effective boson charges and the parameters $\chi_i$ take the same values that in the Hamiltonian (\ref{eq:cqfhamiltonian}).

In the previously defined operators there are a set of free parameters that should be fixed to reproduce as well as possible a set of excitation energies and transition rates, as will be explained in detail in section  \ref{sec-fit-procedure}.

This approach has been successfully used in a set of recent works for Zr \cite{Garc19,Garc20}, Pt \cite{Garc09,Garc11}, Hg \cite{Garc14b,Garc15b} and Po isotopes \cite{Garc15,Garc15c}. 

\subsection{The fitting procedure: energy spectra and absolute $B(E2)$ reduced transition probabilities}
\label{sec-fit-procedure}
In this Section, we present the way the parameters of the Hamiltonian (\ref{eq:ibmhamiltonian}), (\ref{eq:cqfhamiltonian}),  and (\ref{eq:vmix}), as well as the effective charges of the $\hat{T}(E2)$ transition operator (\ref{eq:e2operator}) have been fixed.

We study in this work the even-even isotopes  $^{92-102}$Sr, thereby, covering almost the whole first half of the neutron shell $50-82$. Note that we do not consider the isotopes $^{88-90}$Sr because they are too closed to the neutron shell closure, $50$, and, therefore, the IBM calculations are not reliable enough.

In the fitting procedure carried out here, we try to obtain the best overall agreement with the experimental data including both the excitation energies and the $B(E2)$ reduced transition probabilities. A standard $\chi^2$ is used to obtain the free parameters of the Hamiltonian and the $\hat{T}(E2)$ operator as described in \cite{Garc09,Garc14b,Garc15,Garc19}. Using the expression of the IBM-CM Hamiltonian, as given in equation (\ref{eq:ibmhamiltonian}), and of the $E2$ operator, as given in equation (\ref{eq:e2operator}), in the most general case $13$ parameters show up.  We impose as a constraint that parameters change smoothly in passing from isotope to isotope. Moreover, we try to keep as many parameters as possible at a constant value.
The value of  $\Delta^{N+2}$ evolves, being larger for the lightest isotopes, around $\Delta^{N+2}=1900$ keV, but it drops down to $\Delta^{N+2}=1360$ keV for the majority of isotopes.
\begin{table}
\caption{Hamiltonian and $\hat{T}(E2)$ parameters resulting from the present study.  All quantities have the dimension of energy (given in keV), except $\chi_{N}$ and $\chi_{N+2}$, which are dimensionless and $e_{N}$ and $e_{N+2}$ which are given in units $\sqrt{\mbox{W.u.}}$ }
\label{tab-fit-par-mix}
\begin{center}
\begin{ruledtabular}
\begin{tabular}{cccccccccccccc}
Nucleus&N&$\varepsilon_N$&$\kappa_N$&$\chi_{N}$&$\kappa'_N$&$\varepsilon_{N+2}$& $\kappa_{N+2}$&$\chi_{N+2}$&$\kappa'_{N+2}$& $\omega$& $\Delta^{N+2}$&$e_{N}$&$e_{N+2}$\\
\hline
  $^{92}$Sr & 3&   838&    -32.01&  0.00&  -7.84&     347.2&   -15.00& -0.88&   0.00& 15&   1900 &  1.53&  1.53\footnotemark[1] \\
  $^{94}$Sr & 4&  365&   -50.00&  0.00&   75.01&   451.7&   -41.81&  0.00&   0.00&  15&   1800 &  1.16&   1.53 \\
  $^{96}$Sr & 5&  620&   -35.00&  0.64&   53.43&   242.7&   -20.00&  -0.79&   9.84&  15&   1500 &  1.33&   0.86 \\
  $^{98}$Sr&  6& 526&   -28.19&  1.88&    18.59&   279.1&   -34.96& -0.72&    0.23&  15&    1360 &  0.78&   2.22 \\
  $^{100}$Sr& 7&  526\footnotemark[2]&   -28.19\footnotemark[2]&  1.88\footnotemark[2]&    18.59\footnotemark[2]&   387.3&   -43.16& -0.77&    -2.99&  15&   1360 &  0.78\footnotemark[2]&   1.93 \\
  $^{102}$Sr& 8&  526\footnotemark[2]&   -28.19\footnotemark[2]&  1.88\footnotemark[2]&    18.59\footnotemark[2]&   387.3&   -46.41& -0.77&    -2.99&  15&    1360 &  0.78\footnotemark[2]&   1.93\footnotemark[3]  \\
\end{tabular}
\end{ruledtabular}
\end{center}
\footnotetext[1]{$\hat{T}(E2)$ intruder parameter taken from $^{94}$Sr.}
\footnotetext[2]{Hamiltonian and $\hat{T}(E2)$ parameters for the regular sector taken from $^{98}$Sr.}
\footnotetext[3]{$\hat{T}(E2)$ intruder parameter taken from $^{100}$Sr.}
\end{table}

The resulting values of the parameters for the IBM-CM Hamiltonian and $\hat{T}(E2)$ operator are given in Table \ref{tab-fit-par-mix}. In this table certain parameters could not be determined unequivocally from the experimental information. In particular, the parameters corresponding to the regular sector of $^{100-102}$Sr cannot be fixed because all known experimental states belong to the intruder sector, therefore, we have assumed the same values as for $^{98}$Sr. Moreover, the intruder effective charge for $^{92}$Sr and $^{102}$Sr are taken from $^{94}$Sr and $^{100}$Sr, respectively.
The variation of the parameters from isotope to isotope is relatively smooth, although with some exceptions, as for the effective charge $e_{N+2}$ in $^{96}$Sr, where a sudden drop is observed, without an obvious explanation.
\begin{figure}
  \centering
  \includegraphics[width=0.5\linewidth]{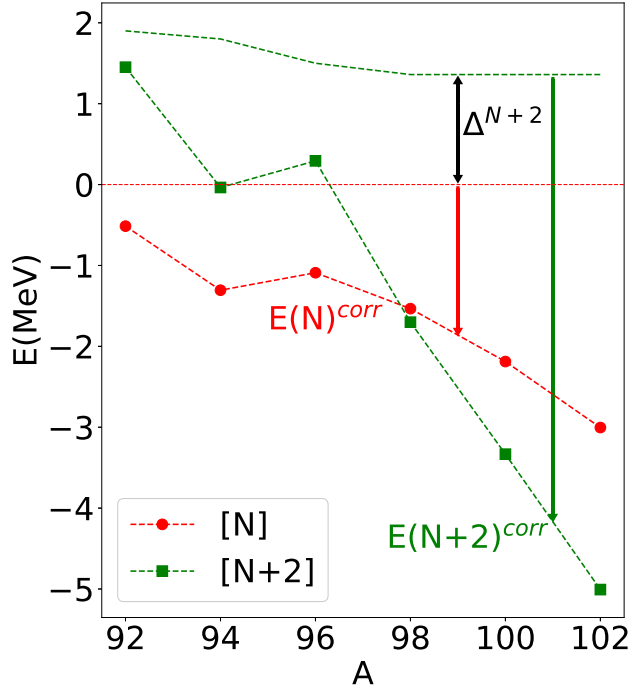}
  \caption{Absolute energy of the lowest unperturbed regular and intruder 0$^+_1$ states for $^{92-102}$Sr. The arrows correspond to the correlation energies in the N and N+2 subspaces (see also the text for a more detailed discussion).}
  \label{fig-energ-corr}
\end{figure}

\section{Correlation energy in the configuration mixing approach} 
\label{sec-corr_energy}
Intruder states are expected to appear, in principle, at much higher energies than the regular ones because a large amount of energy is needed to create a 2p-2h excitation across the shell gap. In the case of Sr, this energy gap corresponds to the proton shell closure at $Z=40$. The energy needed to generate the excitation should be corrected by the pairing energy gain due to the formation of two extra $0^+$ pairs. In our fit, this energy is  $\Delta^{N+2}=1900$ keV for $^{92}$Sr and steady drops till $\Delta^{N+2}=1340$ keV for $^{98-102}$Sr. The former energies do not correspond to the position of the intruder bandhead due to the effect of the so called correlation energy in both the regular and the intruder configurations. Hence, the interaction among the bosons reduces considerably the energy of the configuration, therefore, to know the relative position of regular and intruder states it is needed to understand how behaves this energy gain. First of all, one should consider that this energy gain increases with the number of bosons, being, in principle, larger for the intruder than for the regular configuration. Moreover, it increases as one approaches the mid-shell, where its maximum is.   
\begin{figure}[hbt]
  \centering
  \includegraphics[width=.5\linewidth]{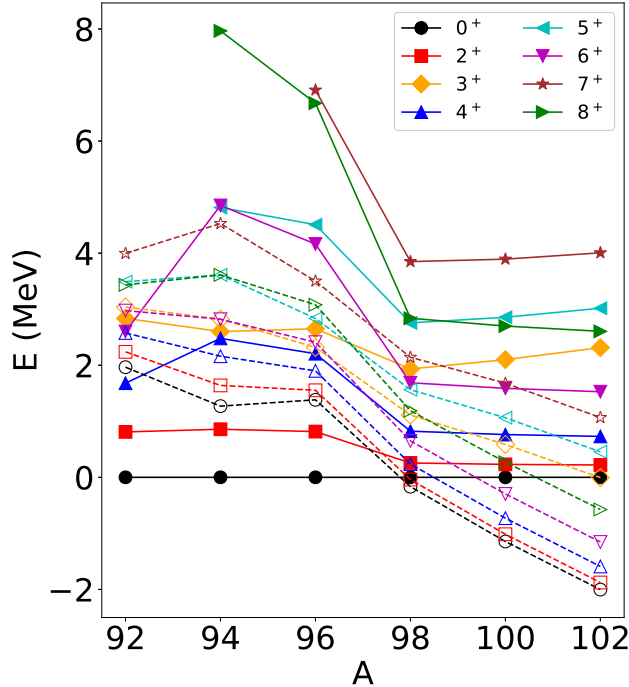}
  \caption{Energy spectra for the IBM-CM Hamiltonian presented in Table \ref{tab-fit-par-mix}, switching off the mixing term. The two lowest-lying regular and intruder states for each of the angular momenta are shown (full lines with closed symbols for the regular states while dashed lines with open symbols are used for the intruder ones).}
  \label{fig-ener-nomix}
\end{figure}

In the framework of the IBM-CM it is quite straightforward to analyze the correlation energy of regular and intruder configurations separately simply switching-off the mixing term between both spaces, obtaining a set of unperturbed energies. In Fig.\ \ref{fig-energ-corr}, we present the value of the reference energy for the regular states, which is set by convention at $0$, and for the intruder ones, which corresponds to $\Delta^{N+2}$. The thick red dashed line represents the unperturbed energy of the regular ground state. As can be seen, there is a relatively large energy gain and it can be understood because of the large value of $|\kappa_N|$. The last two points corresponding to $^{100-102}$Sr should be considered as an extrapolation trend because the value of the regular parameters of the Hamiltonian could not be fixed due to the lack of experimental information. In the case of the intruder configuration, corresponding to the thick green dashed line, the contribution for $^{92}$Sr is quite modest but it steadily increases till reaching a maximum for $^{102}$Sr. The crucial point is that from $^{98}$Sr and onwards, the intruder configuration is below the regular one and the separation in energy between both configurations increases with the neutron number.

In Fig.\ \ref{fig-ener-nomix}, the theoretical spectra of the Sr isotopes with the mixing term fixed to zero and considering as a reference the ground state of the regular configuration is presented. The first worthy aspect is the kink observed in the intruder spectrum for $L>2$,  with a maximum at $A=94$ which could be related to the filling of the $d_{5/2}$ orbit as happens in the case of Zr \cite{Garc20}. The regular spectrum presents a rapid change from a vibrational pattern into a rather rotational one. Note that the regular energy systematics for $A>98$ should be regarded with certain caution because for these nuclei the same parameters as for $^{98}$Sr were taken. In the case of the intruder states, the spectrum shows a quite smooth evolution from a vibrational-like spectrum into a rotational one and, at the same time, its relative position with respect to the regular bandhead is evolving, becoming the ground state of the system from $A=98$ and onwards. 
\begin{figure}[hbt]
  \centering
  \includegraphics[width=0.7\linewidth]{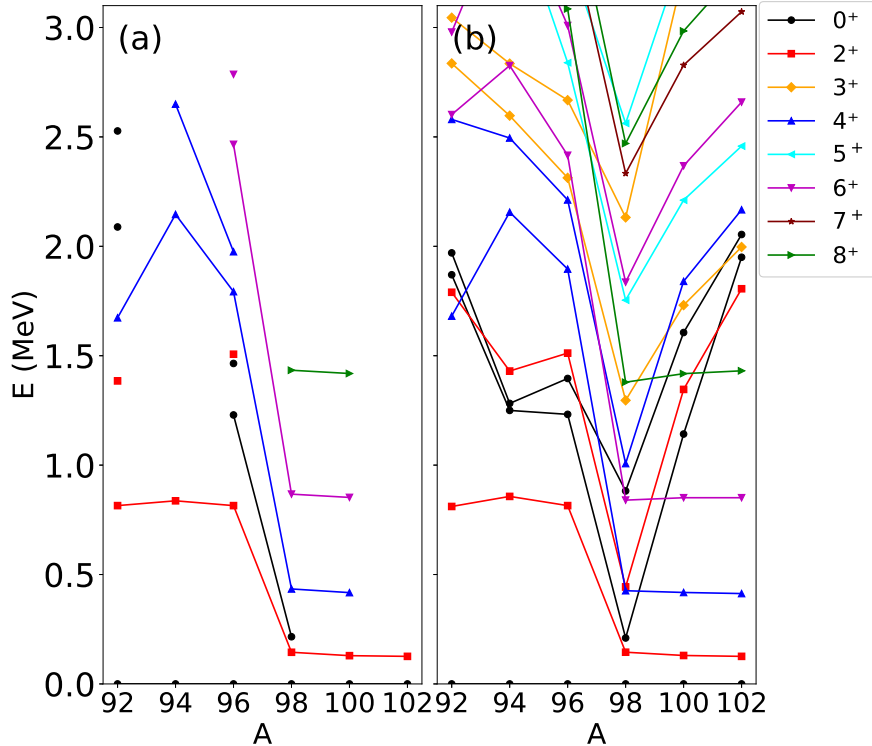}
  \caption{Experimental excitation energies (up to $E \approx 3.0$  MeV) (panel (a)) and theoretical results (panel (b)) obtained from the IBM-CM. 
  Only two excited states (if known experimentally) per angular momentum are plotted.} 
  \label{fig-energ-comp}
\end{figure}

\section{Detailed comparison for energy spectra and $B(E2)$ transition rates}
\label{sec-comp}
\begin{figure}[hbt]
  \centering
  \includegraphics[width=0.6\linewidth]{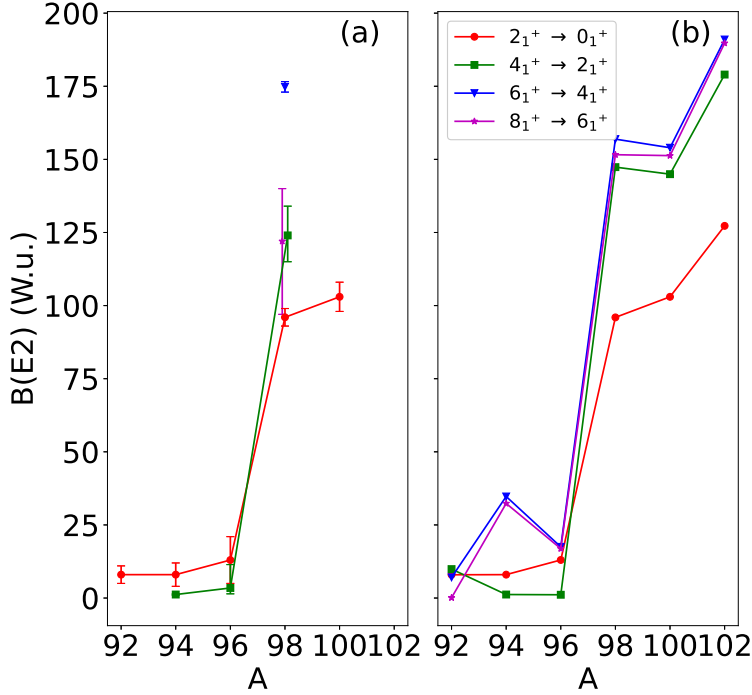}
  \caption{Comparison of the absolute $B(E2)$ transition probabilities along the yrast band, given in W.u. Panel (a) corresponds to known experimental data and panel (b) to the theoretical IBM-CM results. } 
  \label{fig-be2-1}
\end{figure}
In this section, we compare in detail the theoretical calculations with the experimental data up to an energy excitation of $E\approx 3$ MeV. In Fig.\ \ref{fig-energ-comp}(a), we present the experimental excitation energies while in  Fig.\ \ref{fig-energ-comp}(b) the theoretical ones. Note that for the whole chain, the theoretical $2_1^+$ excitation energy closely follows the experimental one because for this level a large weight has been used in the fit procedure (see Section 3.2  of Ref.\ \cite{Garc09} for details). In the case of $A=92$ and $A=94$, the excitation energies are greatly affected by the presence of the neutron shell closure at $50$ and the sub-shell closure at $56$, which produces a kink in the excitation energy of several states, although not so abrupt as in the case of Zr isotopes. The spectrum of $^{92}$Sr has a vibrational-like behavior but the energies of the two-phonon triplet members are quite scattered, while the theoretical ones appear much closed in energy. The experimental information for $^{94}$Sr is quite scarce, with the noticeable absence of experimental excited $0^+$ states (although in \cite{Cruz20} a $0^+$ state at $1880$ keV has been observed) and the kink observed for the $4^+_{1,2}$ states, which is extended up in angular momentum in the theoretical case. Theoretically, the  kink is reproduced and a low-lying excited $0^+$ level is {\it predicted} at an energy similar to the experimental one for $^{96}$Sr. This theoretical value should be regarded with cautious because the calculation has been constrained to get an spectrum consistent with the neighbor nuclei. For $^{96}$Sr, the agreement between theory and experiment is good, except for the states $4_2^+$ and $6_2^+$ which are predicted too high in energy. In the case of $^{98}$Sr, the abrupt change in the structure of the spectrum is perfectly reproduced with a rather rotational spectrum for the yrast band and a very low first excited $0^+$ state. For $^{100-102}$Sr very few experimental levels are known and the theoretical spectra correspond to a rotational yrast band and to a intruder configuration which energy increases as one moves away from the neutron number $60$. 
\begin{figure}[hbt]
  \centering
  \includegraphics[width=0.6\linewidth]{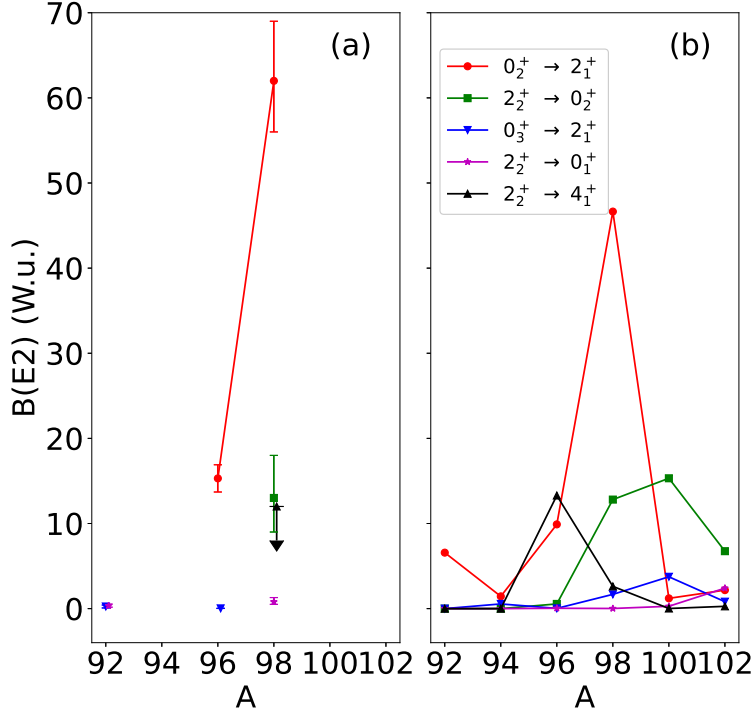}
  \caption{Comparison of the few non-yrast intraband  absolute $B(E2)$ transition  probabilities, given in W.u. Panel (a) corresponds to the known experimental data, panel (b) to the theoretical IBM-CM results. The arrow for $A=98$ represents an upper limit. }
  \label{fig-be2-2}
\end{figure}

The experimental information about $B(E2)$ transition rates is scarce and only few values are know. In Table \ref{tab-be2a}, the known experimental values and the theoretical ones are compared in a detailed way and, moreover, in Figs.\ \ref{fig-be2-1} and \ref{fig-be2-2} intra and interband transitions are depicted, respectively. The agreement between theory and experiment is satisfactory and no noticeable disagreements exist. The appropriated reproduction of the transition rates is a quite stringent test for the model, therefore, the obtained agreement proves that the presented calculations are reliable enough, specially for $^{96}$Sr and  $^{98}$Sr where the experimental information is more abundant.

In Fig.\ \ref{fig-be2-1}, the intraband $B(E2)$ reduced transition probabilities for the yrast band are drawn. The observed trend shows a clear increase in the colectivity as the mass number increases, with an abrupt change when passing from $A=96$ to $A=98$, moreover, showing up the influence of the $d_{5/2}$ filling in the peak that appears for theoretical $6_1^+\rightarrow 4_1^+$ and  $8_1^+\rightarrow 6_1^+$ transitions. In Fig.\ \ref{fig-be2-2}, the interband transitions are depicted. One can see a rather flat behavior for the major part of the transitions, except for $0_2^+\rightarrow 2_1^+$ and $2_2^+\rightarrow 4_1^+$ which present a peak around $A=98$ that is the place where the abrupt change in the nuclear structure of Sr isotopes is known to exist. The transition  $2_2^+\rightarrow 0_2^+$ shows a sudden increase from  $A=98$ and onwards.
\begingroup
\squeezetable
\begin{table}
  \caption{Comparison of the experimental absolute $B(E2)$ values (given in W.u.) with the IBM-CM Hamiltonian results.
    Data are taken from the Nuclear Data Sheets \cite{Bagl12b,Abri06,Negr11,Abri08,Sing03,Chen20,Sing08,Defr09},  
    complemented with references presented in section \ref{sec-exp}.}  
  \label{tab-be2a}
  \vspace{-.5cm}
  \begin{center}
\begin{ruledtabular}
\begin{tabular}{cccc}
Isotope   &Transition             &Experiment&IBM-CM \\
\hline
  $^{92}$Sr&$2_1^+\rightarrow 0_1^+$& 8(3)         & 8\\  
          &$2_2^+\rightarrow 0_1^+$& 0.35(18)       & 0.00 \\   
          &$2_3^+\rightarrow 2_1^+$& $>1.2$       & 0.04 \\   
          &$0_3^+\rightarrow 2_1^+$& 0.25(17)        & 0.28 \\   
\hline
  $^{94}$Sr&$2_1^+\rightarrow 0_1^+$& 8(4)         &8    \\  
          &$4_1^+\rightarrow 2_1^+$& 1.2(2)\footnotemark[1]    & 1.2 \\     
\hline
  $^{96}$Sr&$2_1^+\rightarrow 0_1^+$& 13(8)   & 13    \\  
          &$0_2^+\rightarrow 2_1^+$& 15.3(16)  &  9.90  \\   
          &$0_3^+\rightarrow 2_1^+$& 0.028(11)  &  0.027  \\  
          &$2_2^+\rightarrow 2_1^+$& $>8.9$  &  8.6  \\  
          &$4_1^+\rightarrow 2_1^+$& 3$(^{+8}_{-2})$\footnotemark[1] &  1  \\  
  \hline
  $^{98}$Sr&$2_1^+\rightarrow 0_1^+$& 96(3)         & 96    \\  
          &$0_2^+\rightarrow 2_1^+$& 62$(^{+7}_{-6})$       & 46 \\   
          &$4_1^+\rightarrow 2_1^+$& 124$(^{+10}_{-9})$    & 147   \\  
          &$6_1^+\rightarrow 4_1^+$& 174.8(18)         & 156.9    \\  
          &$2_2^+\rightarrow 4_1^+$& $<12$       & 2.6 \\   
          &$2_2^+\rightarrow 0_2^+$& 13$(^{+5}_{-4})$      & 13   \\  
          &$2_2^+\rightarrow 0_1^+$& 0.8 $(^{+5}_{-3})$        & 0.02    \\  
          &$8_1^+\rightarrow 6_1^+$& 122$(^{+25}_{-18})$       & 152 \\   
          &$10_1^+\rightarrow 8_1^+$& 126$(^{+25}_{-18})$     & 136   \\  
          &$6_2^+\rightarrow 4_1^+$& 0.00010$(^{+4}_{-3})$     & 0.00022   \\    
  \hline
  $^{100}$Sr&$2_1^+\rightarrow 0_1^+$& 103 (5)         & 103    \\   
\end{tabular}
\end{ruledtabular}
\end{center}
\vspace*{-.2cm}
\footnotetext[1]{Data taken from Ref.\ \cite{Regi17}.}
\end{table}
\endgroup

\begin{figure}[hbt]
  \centering
  \includegraphics[width=.85\linewidth]{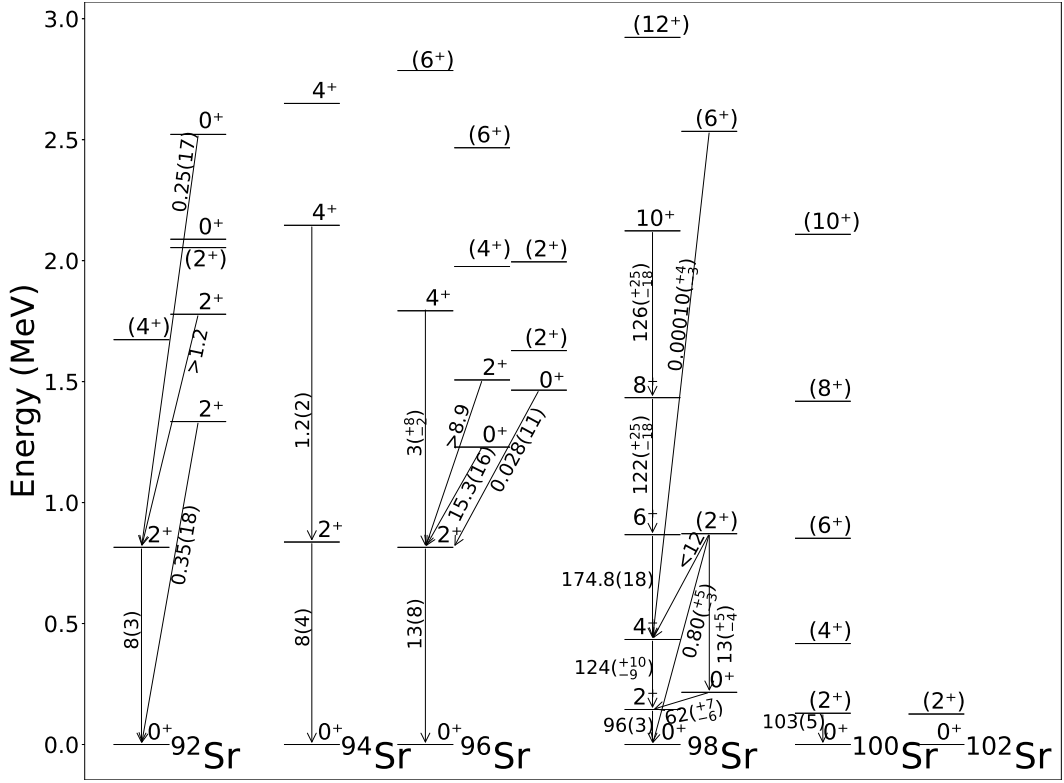}
  \caption{Experimental excitation energies and absolute $B(E2)$ transition rates (given in W.u.) for selected states in $^{92-102}$Sr.} 
  \label{fig-exp-92-102}
\end{figure}

\begin{figure}[hbt]
  \centering
  \includegraphics[width=.85\linewidth]{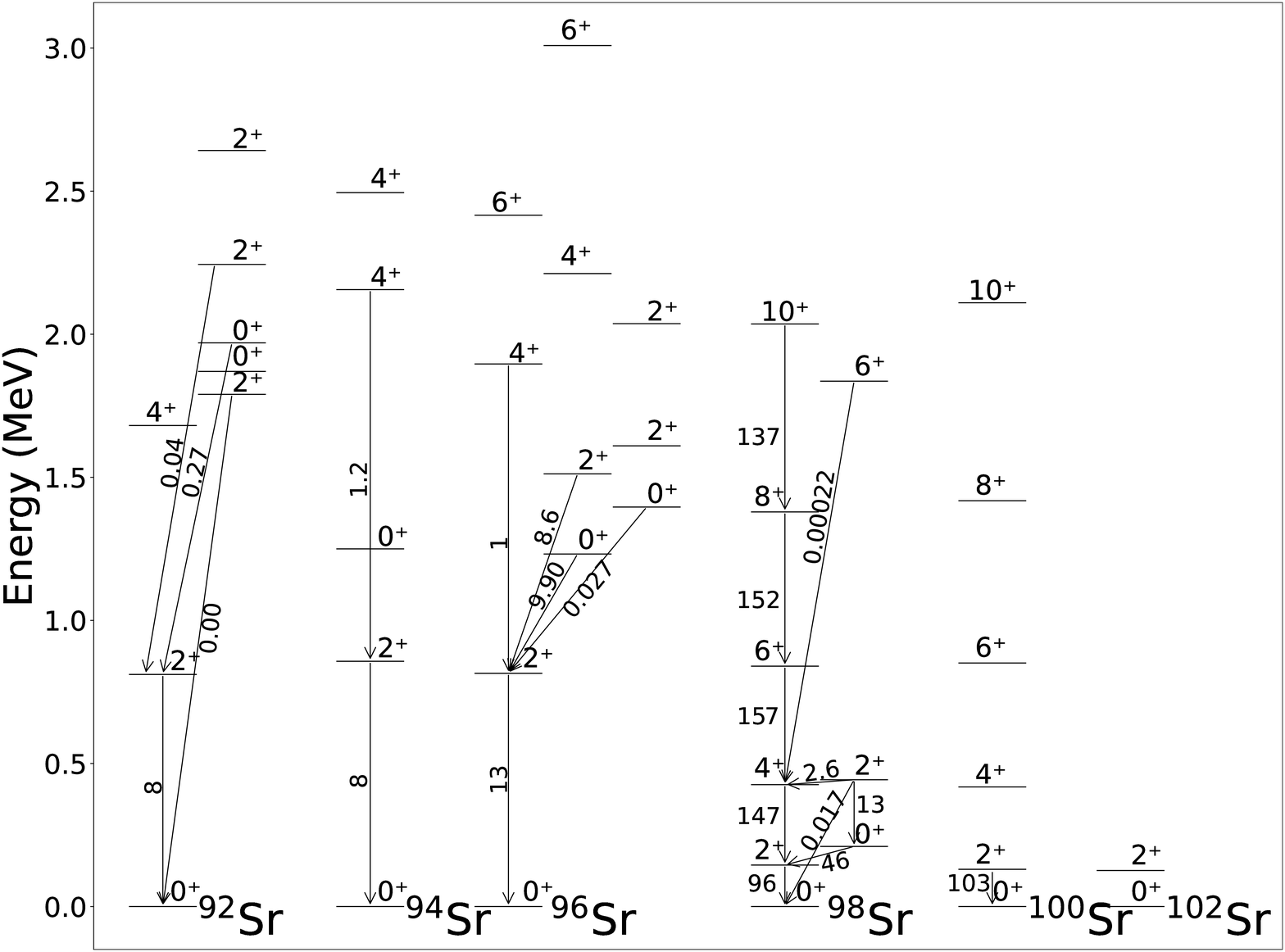} 
  \caption{Theoretical excitation energies and absolute $B(E2)$ transition rates (given in W.u.) for selected states in $^{92-102}$Sr.} 
  \label{fig-theo-92-102}
\end{figure}

In Figs.~\ref{fig-exp-92-102} and \ref{fig-theo-92-102}, the detailed experimental spectra up to an excitation energy $E\approx 3$ MeV and $B(E2)$ transition rates are presented, corresponding to experimental values and theoretical results, respectively. The separation in bands has been carried out first, considering an yrast band and them grouping the rest of levels around a $0^+$ or $2^+$ bandhead. $^{92}$Sr presents a clear vibrational structure with the two-phonon state placed at the double of the energy of the single phonon state, although the energies of the triplet members are quite split, with a very low $B(E2; 2_2^+\rightarrow 0_1^+)$ value connecting the two and the zero phonon state. The rest of states could be identified with intruder states. In the case of $^{94}$Sr, it is hard to disentangle its structure because only 3 excited states below $E_x=3$ MeV are known, moreover, the fact that $B(E2; 2_1^+\rightarrow 0_1^+)>> B(E2; 4_1^+\rightarrow 2_1^+)$  and that $E(4_1^+)>>E(2_1^+)$ hinders the identification as a vibrational nucleus. The theoretical calculation suggests that the $4_1^+$ state has an intruder nature and it also generates two low-lying $0^+$ states, not know experimentally, one intruder and the other of regular nature. The $^{96}$Sr apparently presents a vibrational-like structure but with already a large influence of the intruder states as shown in the departure from the vibrational case of the $B(E2)$ values connecting double and single phonon states. Moreover, the existence of several low-lying $0^+$ states points towards the presence of intruder states. In $^{98}$Sr the yrast band already presents a rotational spacing because the first $0^+$ intruder state becomes the ground state. Concerning the $B(E2)$ values, they present a certain departure from the rotational case for the highest angular momenta. The $0_2^+$ state is the bandhead of the regular band and the observed $B(E2)$ values associated to this band show a clear correspondence with the case of $^{96}$Sr, namely,  $B(E2; 2_2^+\rightarrow 0_2^+)= 13$ W.u. in $^{98}$Sr while  $B(E2; 2_1^+\rightarrow 0_1^+)= 13$ W.u. in $^{96}$Sr too, suggesting the crossing of the regular and the intruder bands. Finally, for $^{100}$Sr  a clear rotational yrast band is observed and for $^{102}$Sr only the state $2_1^+$ is known experimentally.

\section{Wave function structure:  configuration mixing and unperturbed structure}
\label{sec-wf}
\begin{figure}[hbt]
  \centering
  \includegraphics[width=.6\linewidth]{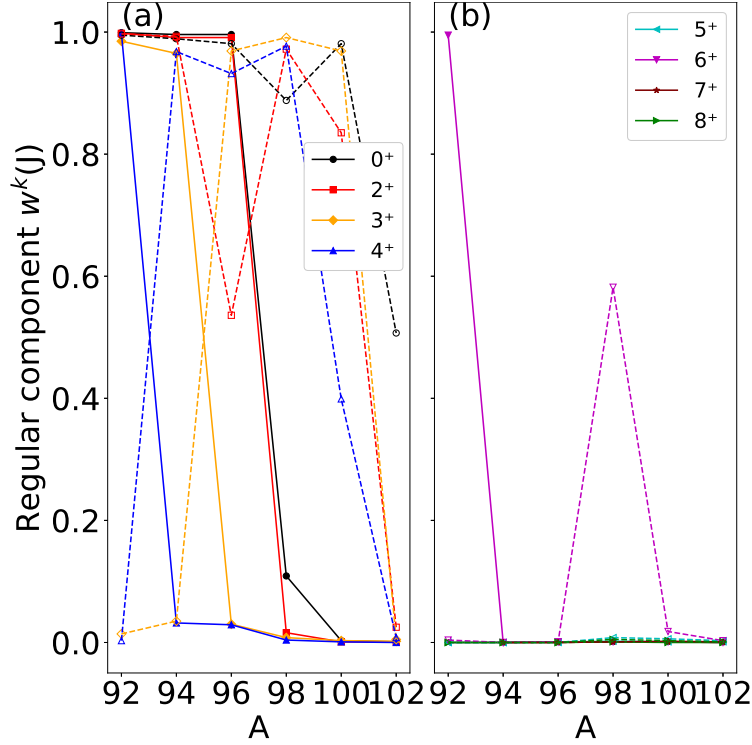}
  \caption{Regular content of the two lowest-lying states
    for each $J$ value (full lines with
    closed symbols correspond with the first state while dashed lines with open
    symbols correspond with the second state) resulting from the IBM-CM
    calculation.}
  \label{fig-wf}
\end{figure}
To disentangle the structure of the states, the first step is to calculate the fraction of the wave function lying in the sectors $[N]$ and $[N+2]$. 
In Fig.\ \ref{fig-wf}, it is drawn the fraction of the wave function within the regular sector, $[N]$, defined as $w^k(J)\equiv \sum_i |a_i^k(J)|^2$ (being the $a$'s the coefficients of the wave function in the regular sector and $k$ an index), for the first two members of each angular momentum (full line for the first and dashed for the second state). The ground state experiences a rapid transition from an almost regular structure, up to $A=96$, into a fully intruder one from $A=98$ and onwards. The same trend is observed for the $2_1^+$, $3_1^+$, $4_1^+$, and $6_1^+$ states, but in this case the transitions appear at $A=98$, $A=96$, $A=94$, and $A=94$, respectively, while for the $5_1^+$, $7_1^+$, and $8^+_1$ states all the time a intruder structure is shown up. 
For the second member of the angular momenta (dashed line), in the case of $5^+_2$, $7^+_2$, and  $8^+_2$, the states are of intruder nature all the way; the $0_2^+$ state has a regular structure except for $A=102$; the $2_2^+$ state is strongly mixed at $A=96$, it is mainly of regular character for $A=92, 94, 98$, and $100$ and of intruder nature for $A=102$; the state $3_2^+$  presents a regular character for $A=96, 98$, and $100$, while intruder for $A=92, 94$, and  $102$; the $4_2^+$ state presents a regular character for $A=94, 96$ and $98$, a intruder one for $A=92$ and $102$ and is mixed for $A=100$; finally the $6_2^+$ state is regular for $A=92$, mixed for $A=98$ and intruder for the rest of isotopes.
\begin{figure}[hbt]
  \centering
  \includegraphics[width=.6\linewidth]{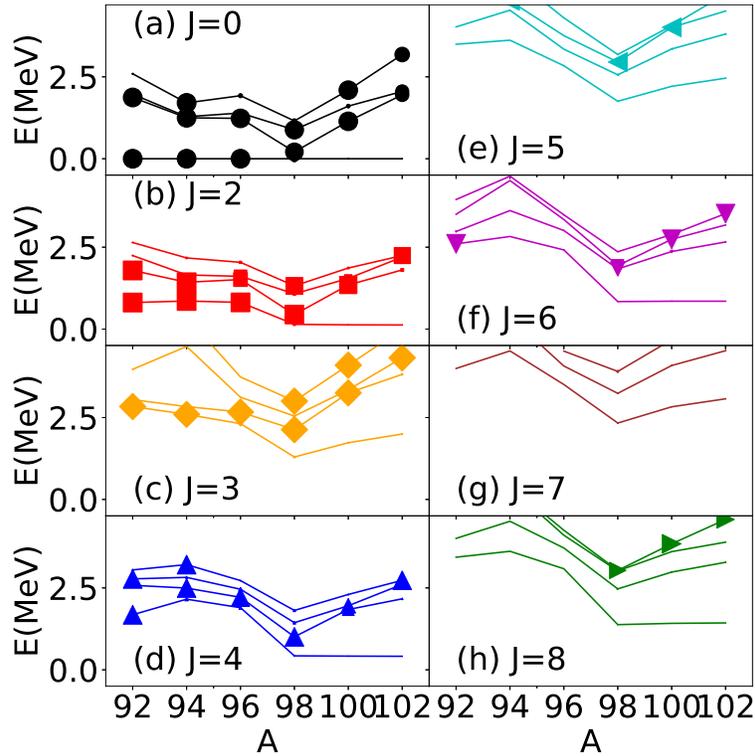}
  \caption{Energy systematics of the four lowest states below $4$ MeV. The size of the symbol is proportional to the value of $w^k(J)$ (see text for details). Each panel corresponds to a given angular momentum, (a) for $J=0$, (b) for $J=2$, (c) for $J=3$, (d) for $J=4$, (e) for $J=5$, (f) for $J=6$, (g) for $J=7$, and (h) for $J=8$.}
  \label{fig-wf-ener}
\end{figure}

The results on the regular content of the wave function that we have presented, so far, in Fig.\ \ref{fig-wf} are strongly influenced by the crossing of the different levels and, therefore, they could mask a smooth trend. To overcome this problem, we present the regular content of the first four states per angular momentum at the same time that the excitation energy of the states, using the size of the dot associated to each state as proportional to the regular content of the wave function. In Fig.\ \ref{fig-wf-ener} we present such information, corresponding the size of the dot for the $0_1^+$ state of $A=92$ to a regular content of $100\%$. For the states $0^+$, $2^+$, $3^+$, and $4^+$ one can see how the regular content ``moves'' from one to another state, depending on the way the levels cross, however for angular momenta $5^+$, $6^+$, $7^+$, and $8^+$ the majority of the states show a intruder character and, therefore, the regular states should appear at much higher energy, not shown in the figure.  

A different decomposition of the wave function that provides extra information is obtained by first calculating the wave functions within the $[N]$,  $|s l,JM\rangle_N$, and $[N+2]$ subspace,  $| l,JM\rangle_{N+2}$, defining an ``intermediate'' basis \cite{helle05,helle08} that corresponds to the states appearing in Fig.\ \ref{fig-ener-nomix}, where the mixing term, $\omega$, has been cancelled. This generates a set of unperturbed bands within the $0p-0h$ and 2p-2h subspaces.
\begin{figure}
\includegraphics[width=.59\textwidth]{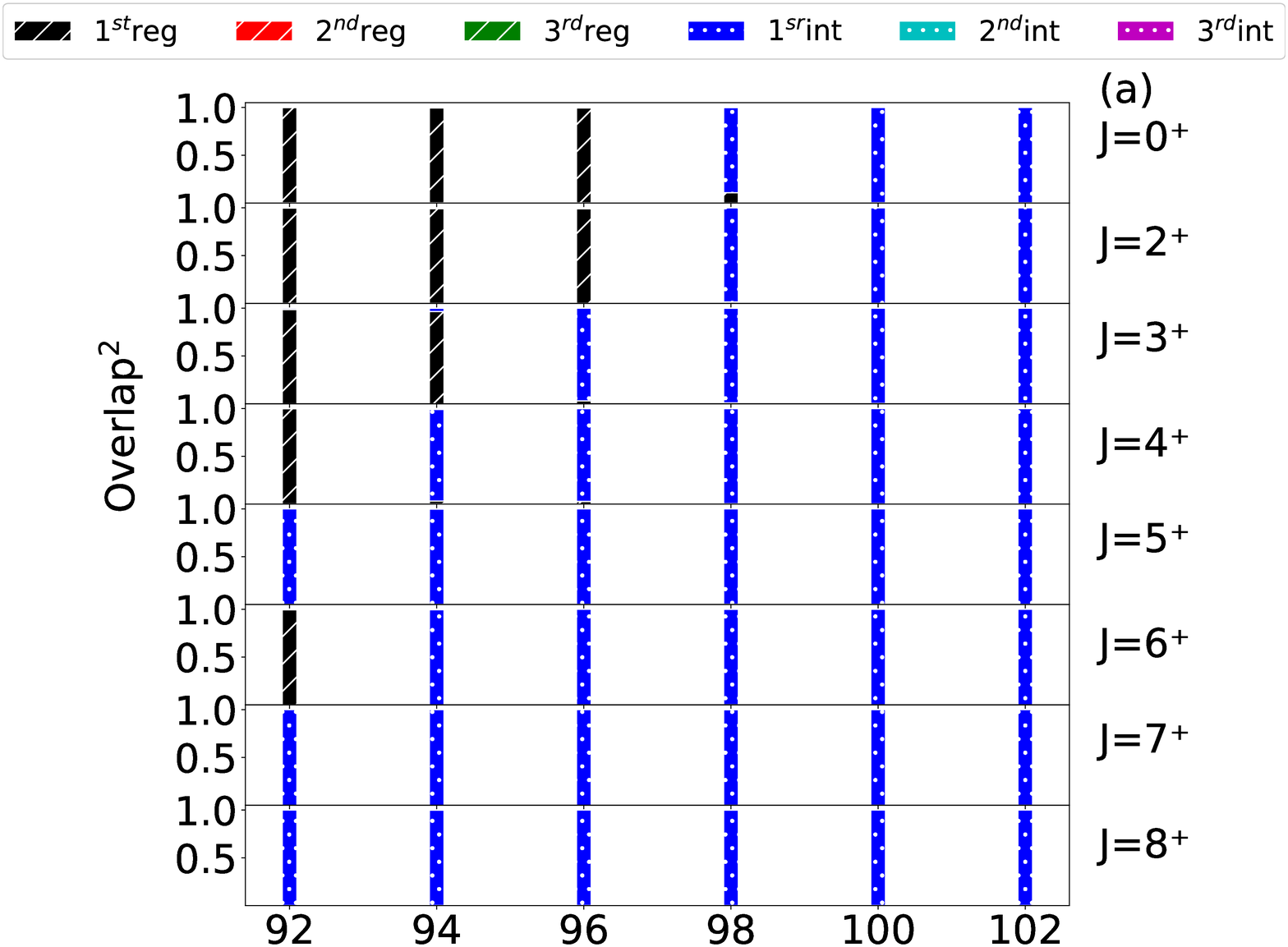}\\
\includegraphics[width=.5\textwidth]{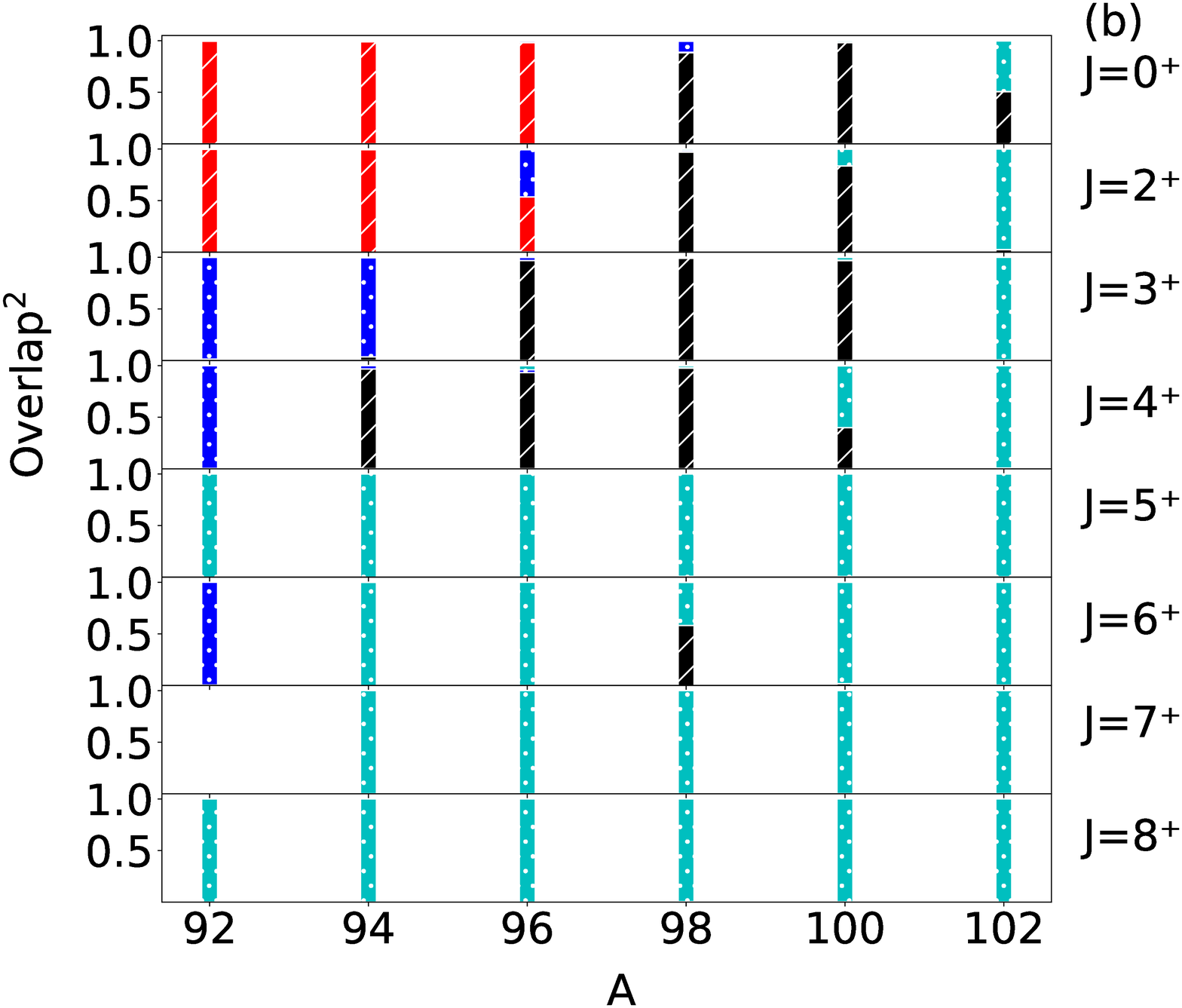}
\caption{Overlap of the wave functions with the wave functions describing the unperturbed basis. Panel (a): overlaps for first $0^+,2^+,3^+,4^+,5^+,6^+,7^+,8^+$ state ($k=1$), panel (b): overlaps for the corresponding second state ($k=2$) (see also text).}
\label{fig-overlap} 
\end{figure} 

The overlaps squared $|_{N}\langle l,JM \mid k,JM\rangle|^2$ and $|_{N+2}\langle m,JM \mid k,JM\rangle|^2$ are depicted in Fig.\ \ref{fig-overlap} for $k=1,2$, $(l,m) =1,2,3$, and  $J^\pi=0^+,2^+,3^+,4^+,5^+,6^+,7^+,8^+$. This way of presenting the wave function shows in a very visual form how the structure of the states evolves along the isotope chain. Regarding the first member of the different angular momenta, for the $0^+$ and $2^+$ states, they correspond to the first regular state, from $A=92$ up to $A=96$, but suddenly change to the first intruder one in the range $A=98-102$; in the case of $3^+$, the first two isotopes correspond to the first regular state while the rest to the first intruder one; for $4^+$ and $6^+$, only $A=92$ fully overlaps with the first regular state and the rest of isotopes overlap with the first intruder one; finally, for $5^+$, $7^+$ and $8^+$, all the states present a complete overlap with the first intruder state. Therefore, the first member of the different angular momenta have a very pure structure without noticeable mixing, presenting a sudden transition when passing from $A=96$ to $A=98$, but moving to lower atomic mass values as $J$ increases. For the second member the reality is much more involve, having the $0^+$ state a complete overlap with the second regular member for $A=92-96$, with the first regular state for $A=98-100$, being a mixture of the first regular and the second intruder state for $A=102$; the state $2^+$ for $A=92-94$ has a full overlap with the second regular state, for $A=96$ it is a mixture of the second regular and the first intruder state, for $A=98-100$, it overlaps with the first regular state, and for $A=102$, with the second intruder one; $3^+$ is a rather pure state corresponding to the first intruder, first regular and second intruder member for $A=92-94$, $A=96-100$, and $A=102$, respectively; $4^+$ is also rather pure except for $A=100$ where a mixture between the first regular and the second intruder state exits, having for $A=92$, $A=94-98$, and $A=102$ a complete overlap with the first intruder, the first regular and the second intruder state, respectively; for $5^+$, $6^+$, $7^+$, and $8^+$, in most of the cases, there is a full overlap with the second intruder state but with some exceptions.       

\section{Study of other observables: radii, isotopic shifts, and two-neutron separation energies}
\label{sec-other}
\begin{figure}[hbt] 
  \centering
  \includegraphics[width=1\linewidth]{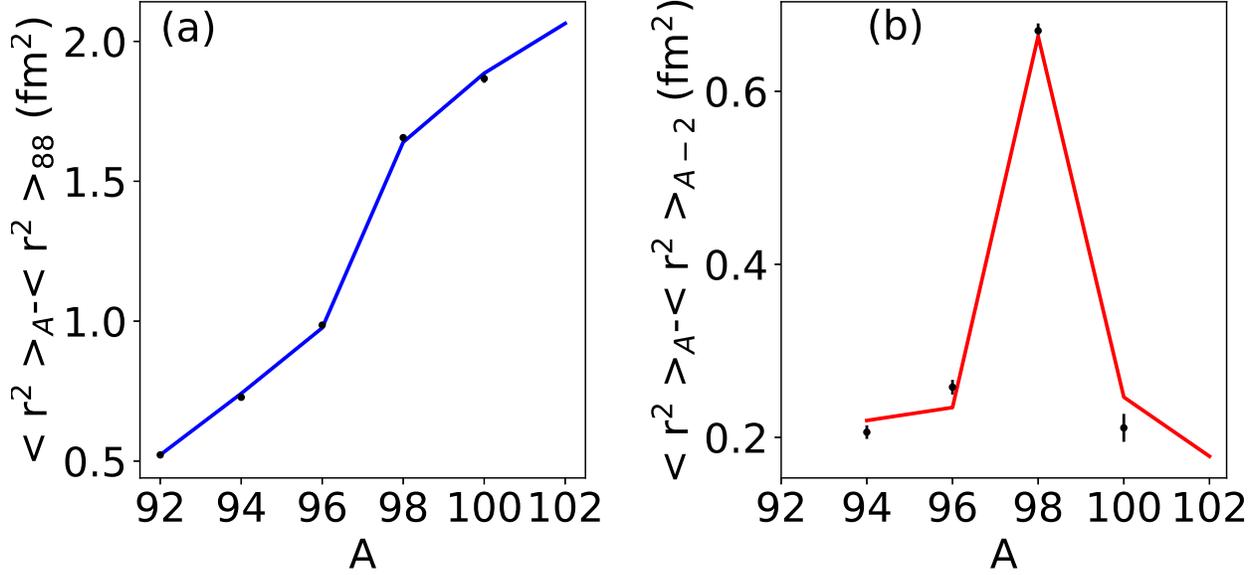}
  \caption{ (a) Charge mean-square radii for the Sr nuclei. (b) Isotopic shift for the Sr nuclei. The data are taken from \cite{Ange13}. Lines for theoretical and dots with error bars for experimental values.
  }
  \label{fig-iso-shift}
\end{figure}

\subsection{Radii and isotopic shifts}
\label{sec-radii}
Nuclear radius is an experimental observable that provides a direct information on the nuclear deformation, hence, it is an excellent probe to detect abrupt changes in nuclear structure, as it is expected to exist at neutron number $60$. In this section we will compare the theoretical value predicted by the IBM-CM with the experimental data \cite{Ange13}.

The value of the radius of a nucleus calculated with the IBM is encoded in the matrix element of the $\hat{n}_d$ operator for the ground state. This value should be superimposed to a linear trend that depends on the number of bosons. Moreover, in the case of the IBM-CM it is also needed to consider both the regular and the intruder configuration. In summary, the radius can be expressed as, \begin{equation}
r^2=r_c^2+ \hat{P}^{\dag}_{N}(\gamma_N \hat{N}+ \beta_N
\hat{n}_d)\hat{P}_{N} + 
\hat{P}^{\dag}_{N+2}(\gamma_{N+2} \hat{N}+ \beta_{N+2} \hat{n}_d) \hat{P}_{N+2},
\label{ibm-r2}
\end{equation}
where $\hat{P}$ are projection operators, the appearing parameters are common for the whole chain of isotopes and they are fixed to reproduce as well as possible the experimental data, which are referred to $^{88}$Sr. $r^2_c$ is choose to match the radius of $^{92}$Sr. The resulting values are  $\gamma_N=0.23$ fm$^2$, $\beta_N=-0.028$ fm$^2$,  $\gamma_{N+2}=0.294$ fm$^2$, and  $\beta_{N+2}=-0.16$ fm$^2$. A similar approach is conducted in \cite{Zerg12} but only considering a single configuration.

The overall description of the radii and even of the isotope shifts is rather satisfactory, as can be readily observed in Fig.\ \ref{fig-iso-shift}, reproducing nicely the sudden onset of deformation in $^{98}$Sr. This feature points towards the appropriated reproduction of the crossing of two configurations with a rather different deformation which is, as a matter of fact, responsible for the abrupt increase of the nuclear radius at $A=98$. 

\subsection{Two-neutron separation energies}
\label{sec-s2n}
\begin{figure}[hbt]
\centering
\includegraphics[width=0.50\linewidth]{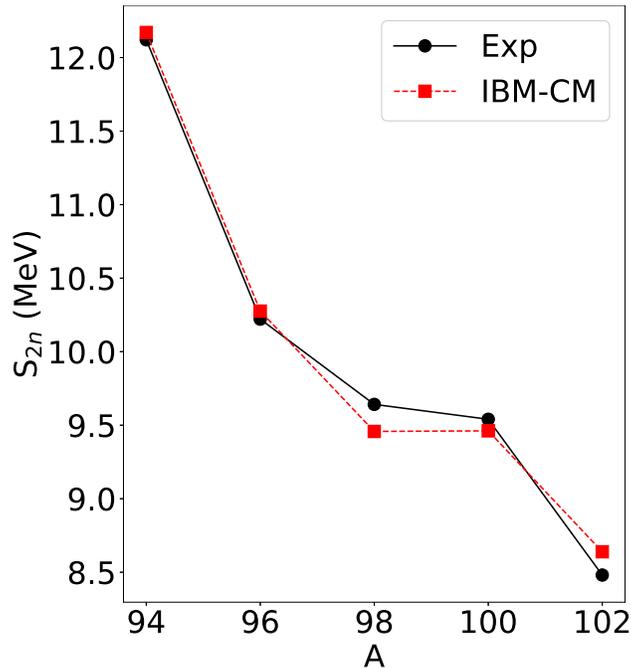}
\caption{Comparison of experimental and theoretical two-neutron separation energies.}
\label{fig-s2n}
\end{figure}
The definition of the S$_{2n}$ involves the value of the binding energy of two neighbouring nuclei separated by two units of mass, as written in this equation,
\begin{equation}
\label{s2n}
S_{2n}(A)=BE(A)-BE(A-2). 
\end{equation}
where $BE$ denotes the binding energy of the nucleus. In the case of the IBM, it is needed to add to the obtained binding energy a contribution that depends on $N$ and $N^2$, not affecting the excitation energies. This leads to an extra linear contribution to the value of the S$_{2n}$ \cite{Foss02}. Therefore, the S$_{2n}$ can be written as,
\begin{equation}
S_{2n}(A)={\cal A} +{\cal B} A+BE^{lo}(A)-BE^{lo}(A-2), 
\label{s2n-lin}
\end{equation}
where $BE^{lo}$ is the local binding energy derived from the IBM Hamiltonian, and the coefficients ${\cal A}$ and ${\cal B}$ are assumed to be constant for a chain of isotopes \cite{Foss02}. In the case of IBM-CM calculations we expect that the effective number of bosons for the ground state can be affected by the influence of the intruder states. To consider this effect, we propose as an \textit{ansazt}, 
\begin{equation}
S_{2n}(A)={\cal A} +{\cal B} (A+2(1-w))+BE^{lo}(A)-BE^{lo}(A-1),
\label{s2n-lin-new}
\end{equation}
where $\omega=\omega^1(0)$ ($w^k(J)\equiv \sum_i |a_i^k(J)|^2$). The values of ${\cal A}$ and ${\cal B}$ are determined, once the $BE^{lo}$'s are known, through a least-squares fit to the experimental values of S$_{2n}$, as explained in detail in \cite{Foss02,Garc14b,Garc19}. In our case, the obtained values are ${\cal A}=52.9$ MeV and ${\cal B}=-0.441$ MeV  and in Fig.\ \ref{fig-s2n} the comparison between experiment and theory is shown, being remarkable the good agreement around neutron number $60$ $(A=98)$ where the flattening of the curve is more evident, corresponding to the place where regular and intruder configurations cross. This type of behavior could be identified with the onset of a quantum phase transition (QPT) \cite{Sach11}, as it was already suggested for the case of Zr \cite{Garc20}.

\section{Nuclear deformation and mean-field energy surfaces}
\label{sec-deformation}
The goal of this section is to obtain information about the evolution of the deformation along the isotope chain, taking into account that deformation ($\beta$) is not a direct observable. The use of different methods to determine the deformation will allow us to analyze the consistency of the obtained values. 
\begin{figure}
\includegraphics[width=.9\textwidth]{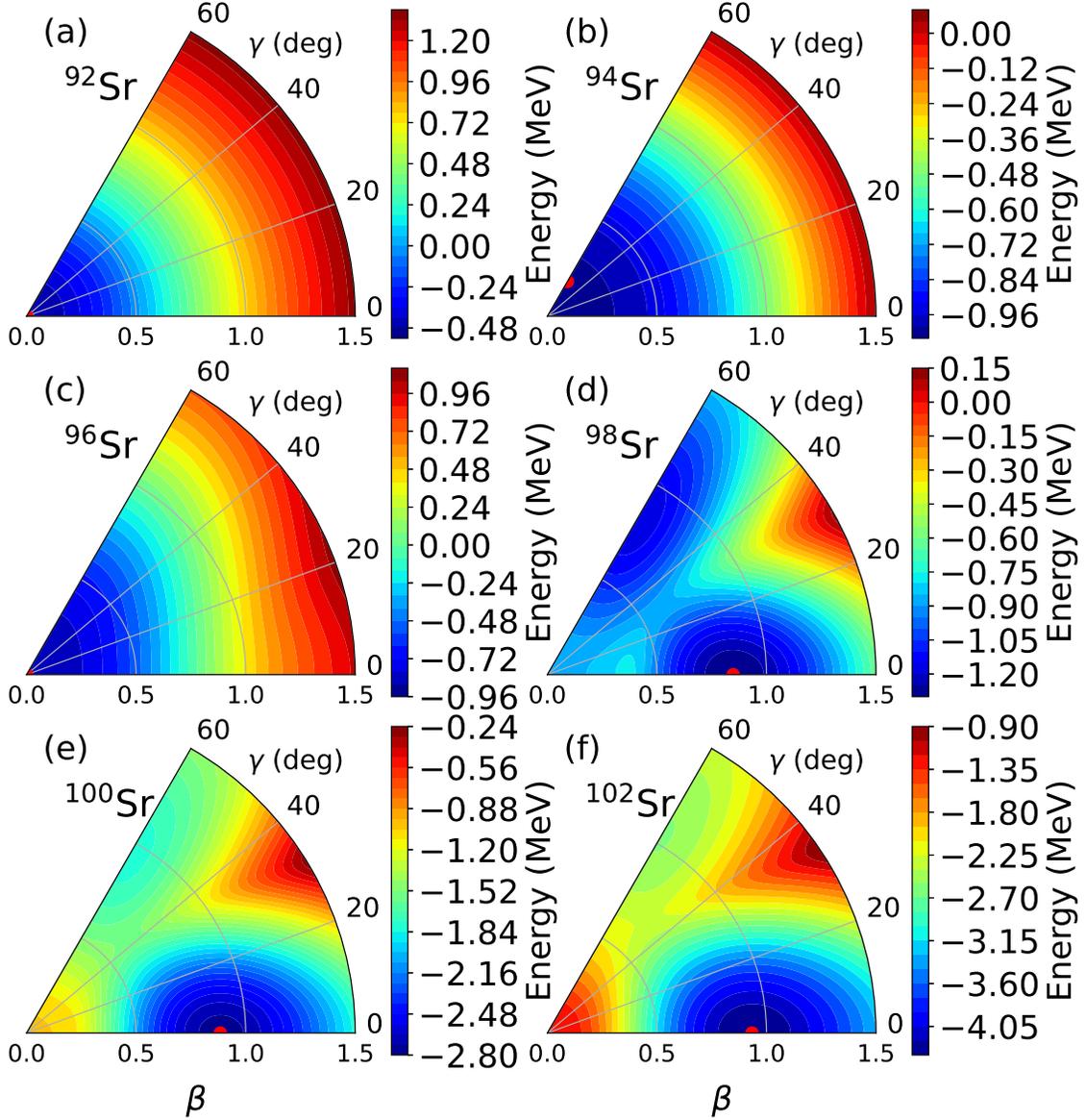}
\caption{Matrix coherent-state calculation for $^{92-102}$Sr, corresponding with the present IBM-CM Hamiltonian (table \ref{tab-fit-par-mix}). They are plotted in the $(\beta, \gamma)$ plane in the range $0.0 \leq \beta \leq 1.5 $ and $0\textrm{(deg)}\leq \gamma \leq 60 \textrm{(deg)}$. The red dot marks the position of the absolute minimum.}
\label{fig_ibm_ener_surph}
\end{figure}

The first way to get information about deformation is through the use of the intrinsic state formalism of the IBM. This formalism was presented in the set of seminal works \cite{gino80,diep80a,diep80b,Gilm74} and it provides a geometric interpretation of the model, defining, in particular, a deformation parameter. This intrinsic state formalism cannot be used directly for the IBM-CM, but the formalism should be enlarged. Frank {\it et al.}\, introduced almost twenty years ago the matrix coherent-state method  \cite{Frank02,Frank04,Frank06,Mora08} which allows to calculate the total energy surface for a system where regular and intruder configurations exist. In Refs.~\cite{Garc14a,Garc14b,Garc15,Garc19}, a detailed description of the method and its application to Pt, Hg, Po, and Zr isotopes, respectively, has been given. In Fig.\ \ref{fig_ibm_ener_surph}, the IBM-CM mean-field energy surfaces are presented for the isotopes $^{92-102}$Sr. In these nuclei, one can see how the three lightest isotopes present a rather spherical shape, although with a small deformation for $^{94}$Sr and a rather flat well for $^{96}$Sr, however, the three heaviest ones are well deformed with a prolate shape, presenting $^{98}$Sr a clear coexistence between a prolate and an oblate minima being almost degenerate. In $^{100-102}$Sr, the nuclei present a well deformed prolate minimum coexisting with an oblate one at a higher energy. The obtained energy surfaces are in close correspondence with the ones appearing in \cite{Rodr10,Nomu16} where a HFB calculation using a Gogny interaction is performed. In that work, $^{92}$Sr  is still spherical and a small oblate deformation already appears for $^{94}$Sr, in $^{96}$Sr an oblate and a prolate minimum already coexist, corresponding the deepest minimum to the oblate shape and the same holds for $^{98-102}$Sr but, in these cases, the prolate minimum is the deepest one. It is worth to mention that in Zr isotopes a quite similar situation exists but with the coexistence of a spherical and a prolate minimum (at the IBM-CM energy surface) while oblate and prolate in the case of Sr. 
\begin{figure}[hbt]
\centering
\includegraphics[width=0.50\linewidth]{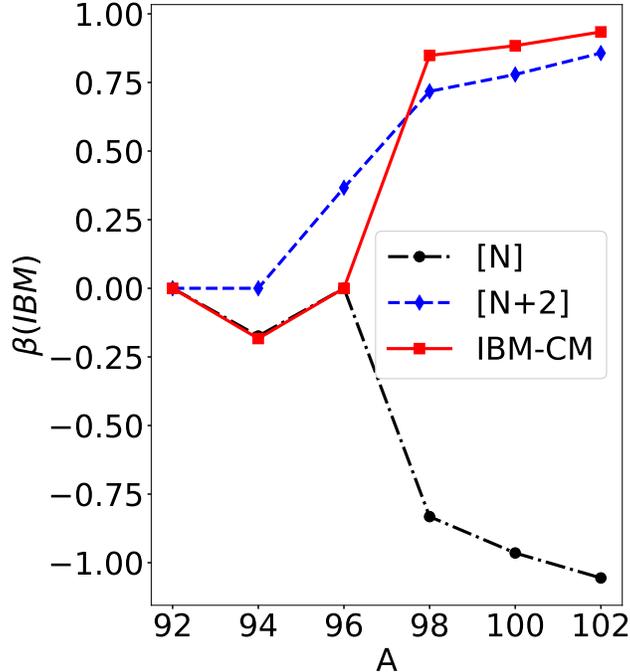}
\caption{Value of $\beta$ extracted from the IBM-CM energy surface. $[N]$, $[N+2]$, and IBM-CM correspond to a pure regular configuration, a pure intruder configuration, and to the complete calculation.}
\label{fig-beta}
\end{figure}

To better illustrate the coexistence of two different shapes, in Fig.\ \ref{fig-beta} we depict the value of $\beta$ corresponding to the absolute minimum resulting from the present IBM calculation. Note that this IBM $\beta$ variable is not equivalent to the one of the collective model but can be approximately connected as explained in \cite{gino80}.  On the one hand, we present the values for the regular and the intruder sector separately, labeled with $[N]$ and $[N+2]$, respectively, assuming that no interaction exits between them, imposing $\omega=0$, and, on the other hand, we present the value of $\beta$ obtained from the full calculation, IBM-CM. In this figure, it is readily appreciated how both configurations are evolving from an almost spherical shape into an oblate ($[N]$) or a prolate one ($[N+2]$), being a little more abrupt the onset of deformation in the case of the regular configuration than in the intruder case. The complete calculation shows how the first three isotopes correspond to the regular configuration (spherical), while the last three isotopes to the intruder one (prolate).  

Even though  the shape of a certain nucleus is not an observable, the study of Coulomb excitation allows us to extract, in an almost model independent way, information about nuclear deformation, as shown by Kumar, Cline and coworkers \cite{kumar72,Cline86}. This approach is based on the use of the concept of ``equivalent ellipsoid'' of a given nucleus, corresponding to an ellipsoid defined as uniformly charged with the same charge, $\left\langle r^{2}\right\rangle$, and $E2$ moments as the original nucleus characterized by a specific eigenstate.  Starting from measured data on various transitions using Coulomb excitation methods, it turns out that the data allow to extract for a given state with defined angular momentum an extra set of two numbers that fit with the variables of the collective model, $\beta$ and $\gamma$. More recently, in Refs.\ \cite{Alha14,Gilb18,Must18} the authors have addressed the question of nuclear deformation within the
laboratory frame, even extending toward excited states making use of the auxiliary-field Monte Carlo method.
\begin{table}
\caption{Values of deformation, $q^2$ and $\gamma$, extracted from the quadrupole shape invariants, together with the value of $w^k$ ([N] content).}
\label{tab-q-invariant}
  \begin{center}
\begin{tabular}{ccccc}
\hline
A & State & $q^{2}$ ($e^{2} b^{2}$) & ~~$\gamma$~~ & $w^k$\\
\hline
92 & 0$_{\text{1}}^+$ & 0.10 & 29 & 0.999\\
  & 0$_{\text{2}}^+$ & 0.06 & 17 & 0.995\\ 
  & 0$_{\text{3}}^+$ & 1.18& 19 & 0.003\\  
  \hline                                  
94 & 0$_{\text{1}}^+$ & 0.10 & 30 & 0.996\\ 
  & 0$_{\text{2}}^+$ & 0.05 & 30 & 0.989\\   
  & 0$_{\text{3}}^+$ & 0.32 & 30 & 0.013\\ 
  \hline                                  
96 & 0$_{\text{1}}^+$ & 0.17 & 39 & 0.996\\
  & 0$_{\text{2}}^+$ & 0.14 & 49 & 0.981\\      
  & 0$_{\text{3}}^+$ & 0.16 & 15 & 0.020\\
  \hline                                  
 98 & 0$_{\text{1}}^+$ & 1.33 & 10 & 0.109\\                                               
   & 0$_{\text{2}}^+$ & 0.31 & 15 & 0.888  \\    
  & 0$_{\text{3}}^+$ & 0.14 & 27 & 0.947  \\     
   \hline                                  
 100 & 0$_{\text{1}}$ & 1.46 & 10 & 0.109  \\    
   & 0$_{\text{2}}^+$ & 0.25 & 48 & 0.981  \\    
   & 0$_{\text{3}}^+$ & 0.98 & 18 & 0.039 \\     
   \hline                                                                                
 102 & 0$_{\text{1}}^+$ & 1.84 & 9 & 0.001 \\    
      & 0$_{\text{2}}^+$ & 0.81 & 19 & 0.507  \\ 
      & 0$_{\text{3}}^+$ & 0.83 & 18 & 0.493 \\  
  \hline

\hline                                      
\end{tabular}
\end{center}
\end{table}

In order to obtain, from a theoretical point of view, the nuclear shape, we will make use of the quadrupole shape invariants. Hence, the quadrupole deformation for the $0^+$ states corresponds to
\begin{eqnarray}
q_{2, i} &=&\sqrt{5}\left\langle 0_{i}^{+}\left|[\hat{Q} \times \hat{Q}]^{(0)}\right| 0_{i}^{+}\right\rangle, \label{eq11} \\
q_{3, i} &=&-\sqrt{\frac{35}{2}}\left\langle 0_{i}^{+}\mid[\hat{Q} \times \hat{Q} \times \hat{Q}]^{(0)} \mid 0_{i}^{+}\right\rangle. \label{eq12}
\end{eqnarray}
The deformation parameters are directly connected with the ones of the triaxial rigid rotor, $q$ and $\delta$, accordingly to, 
\begin{equation}
q=\sqrt{q_{2}}, 
\label{eq15}
\end{equation}

\begin{equation}
\delta=\frac{60}{\pi} \arccos \frac{q_{3}}{q_{2}^{3 / 2}},  
\label{eq16}
\end{equation}
where $\delta$ coincides with the parameter $\gamma$ of the Bohr-Mottelson model up to first order approximation. It is worth to mention the work \cite{Pove20} where a method to calculate the fluctuations of $\beta$ and $\gamma$ is presented.

The theoretical values of $\gamma$ and $q^2$ are presented in Table \ref{tab-q-invariant} simultaneously with the fraction of wave function belonging to the regular sector $w^k$, for every $0_1^+$, $0_2^+$ and $0_3^+$ of the whole chain of Sr isotopes. According to this table, one readily observe the coexistence of different deformations, typically less deformed for the regular than for the intruder states. Another notable feature is the fact that the nature of the deformation of the ground state suddenly increase when passing from $^{92-96}$Sr to  $^{98-102}$Sr. Concerning triaxiallity, the three first isotopes present clear triaxial shapes, while in the three last ones the ground state is close to a prolate shape and the rest of states are rather triaxial.   
\begin{figure}[hbt]
  \centering
  \includegraphics[width=.45\linewidth]{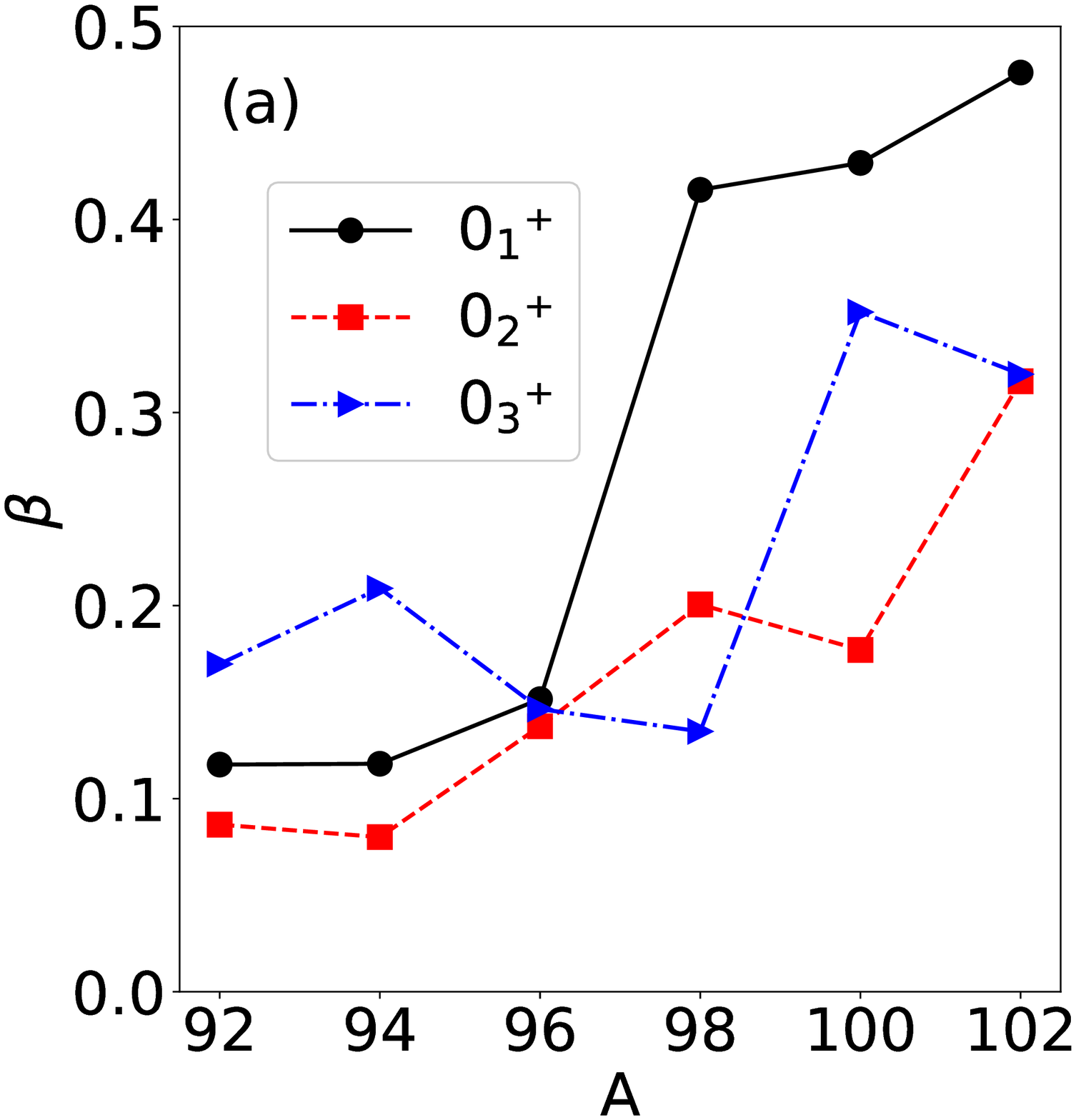}
  ~~
  \includegraphics[width=.45\linewidth]{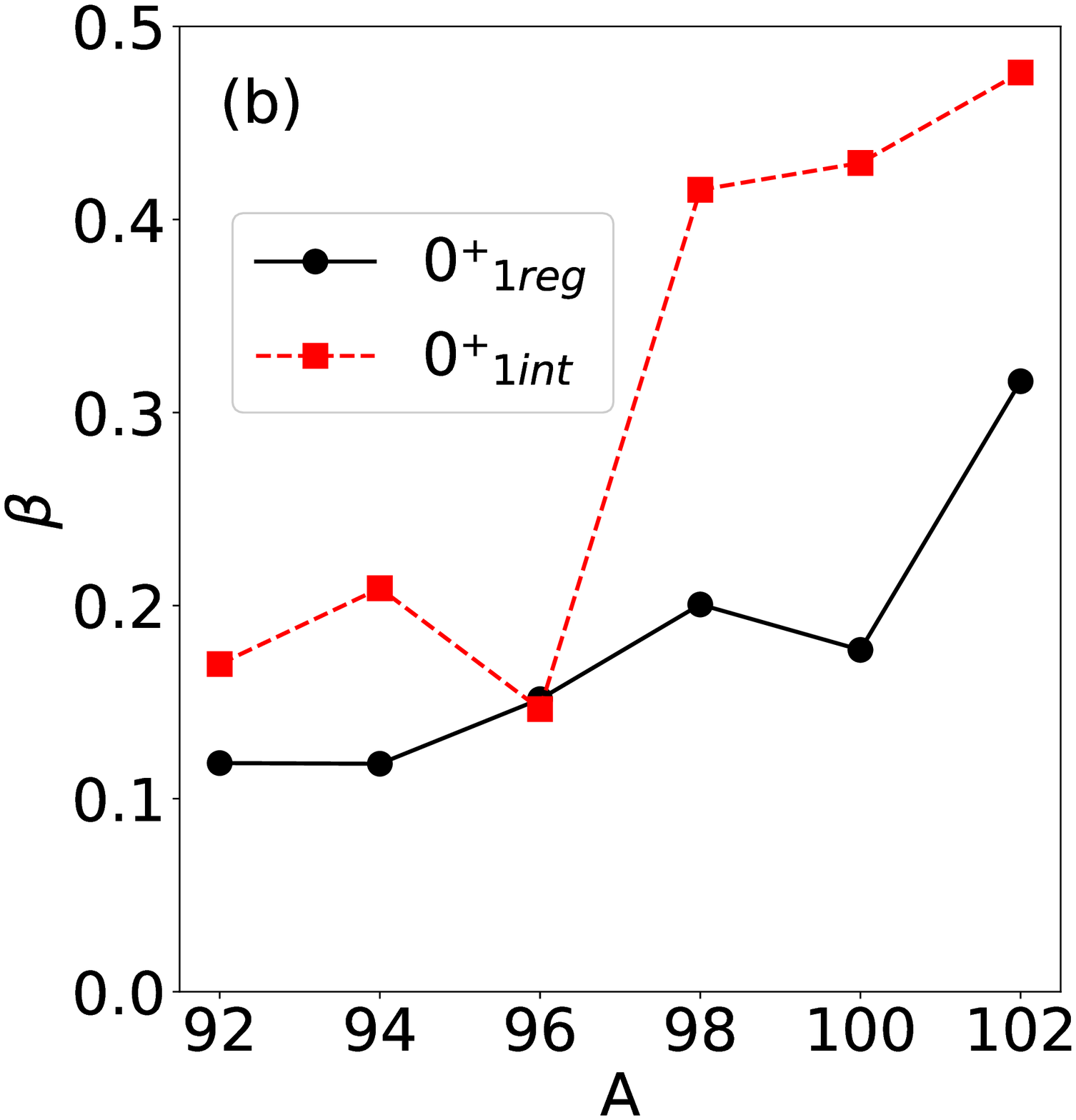}
  \caption{Value of the deformation, $\beta$, extracted from the value of the quadrupole shape invariants. In (a) we plot the value for the first three $0^+$ states. In (b) we plot the same but for the first regular and the first intruder $0^+$ state (see text).}
  \label{fig-beta-fromQ2}
\end{figure}

The deformation parameter $\beta$ can also be obtained from the quadrupole shape invariant (\ref{eq11}) (see, e.g., references \cite{Srebrny06,clement07,Wrzo12}),
\begin{equation}
\beta = \dfrac{4\pi\sqrt{q_2}}{3 Z e r_0^2 A^{2/3}},
\label{beta}
\end{equation}
where $e$ is the proton charge and $r_0 = 1.2$ fm.

The values of $\beta$ extracted from Eq.\ (\ref{beta}) for the corresponding ground state, $0_1^+$, and the $0_2^+$ and $0_3^+$ states, are shown in Fig.\ \ref{fig-beta-fromQ2}(a), where $\beta$ for the ground state presents a rapid increase, passing from a value around $0.15$ for $^{92-96}$Sr to another around $0.45$ for $^{98-102}$Sr. The value of $\beta$ for the $0_2^+$ state steadily increase while for the $0_3^+$ state there also exists a sudden increase for $^{100-102}$Sr. It is very enlightening to present the value of the deformation for the first $0^+$ state belonging to the regular sector, $0^+_{reg}$, and the one of the first $0^+$ state which belongs to the intruder sector, $0^+_{int}$, as shown in Fig.\ \ref{fig-beta-fromQ2}(b) which is done simply taking into account the regular content of the states presented in the last column of Table \ref{tab-q-invariant}. The $0_{reg}^+$ state shows a smooth increase of deformation over the whole chain of isotopes and it corresponds to the $0_1^+$ state for $^{92-96}$Sr and to the $0^+_2$ for $^{98-102}$Sr, while the state $0_{int}^+$ shows the sudden increase of the deformation previously explained and it corresponds to the states $0_3^+$ and $0_1^+$ for $^{92-96}$Sr and $^{98-102}$Sr, respectively.  

As a matter of conclusion of this section, we can say that both approaches presented in this section provide a consistent view of the deformation of the two families of states, intruder and regular, with both configurations close to a spherical shape in $^{92-96}$Sr and an increase in deformation when entering in the region of $^{98-102}$Sr. The mean-field energy surfaces point to the coexistence of a prolate and an oblate shape in $^{98-102}$Sr, corresponding the deepest minimum to the prolate shape. The abrupt onset of deformation, specially for the ground state, which is the result of the crossing of the regular and the intruder configurations, points to the existence of a QPT around $^{98}$Sr. The behavior of the S$_{2n}$ and the nuclear radii, presented in previous section, point to the existence of a QPT, as well. This is a case that resembles the situation in even-even Zr isotopes where also two configurations coexist and they cross at $^{100}$Zr. Therefore, the study of both chain of isotopes suggests the presence of a first order QPT at neutron number $60$ associated to the crossing of two configurations. 

\section{Conclusions and outlook}
\label{sec-conclu}
In this work we have conducted an IBM-CM calculation for the even-even $^{92-102}$Sr. The parameters of the Hamiltonian have been fixed through a least-squares procedure to the excitation energies of states below $3$ MeV and to the corresponding B(E2) values. The number of known experimental B(E2) values is quite reduced in most of isotopes and even the number of experimental excitation energies in some cases. This fact makes impossible to fit certain parameters in some isotopes and in this situation the parameters are taken from the ones of the neighbour isotopes. Once the parameters have been determined, experimental data and theoretical values are compared, showing a more than reasonable agreement, reproducing the rapid changes observed around $A=98$, where the systematics of the density of energy levels suddenly increase and the B(E2) values rapidly increase, as well. The calculation of unperturbed bands, without the mixing term, clearly shows the crossing of two families of states with a much larger gaining of correlation energy in the case of the intruder configuration from $^{98}$Sr and onwards. The wave functions have been also studied in detail clearly showing abrupt changes around $A=98$, passing the low-lying states from being mostly of regular character to being of intruder nature. The two-neutron separation energies and the value of the radii have been computed and compared with the experimental information, presenting a rather good agreement, reproducing the rapid changes that characterize the nuclei around $^{98}$Sr. This agreement is specially relevant because the structure of the states is fixed from the Hamiltonian parameters and, therefore, the correct reproduction of this experimental information points towards a reasonable description of the wave functions. Finally, the mean-field energy surfaces and the value of the deformation have been obtained and a remarkable agreement with previous mean-field calculations is observed.


The analysis of Sr isotopes complements the previous works carried out for Zr isotopes \cite{Garc19,Garc20} where a more detailed data basis has been available. The obtained results for both families of isotopes draw a common landscape where two families of regular and intruder states cross around neutron number $60$ inducing characteristic features, such as the lowering of $0^+$ states, the compression of the spectrum, the enhancement of $E2$ transition rates, the flattening of the two-neutron separation energy or the sudden increase of the nuclear radius. One can consider the later features as hints of the onset of deformation that in our case is generated by the crossing of two configurations with different degree of deformation. 

The obtained regular and intruder configurations in Sr isotopes present a reduced interaction between them, which is at the origin of the observed abrupt changes. There are other isotope chains where regular and intruder configurations also cross for the ground state, as in Pt and Po, but in these cases, the interaction between both families of states is so large that all abrupt changes have been smoothed out, however, when the unperturbed configurations are studied, the results are very similar to the obtained in this work. 


Further experimental information for Sr isotopes will allow to better determine some of the parameters of the Hamiltonian and the $E2$ transition operator. 
As an outlook,  we hope that from the experimental side, effort is enhanced to expand the present data basis with measurements of  excitation energies and both $E2$ and $E0$ transition rates which are well known to give essential information on the variation of the shape associated with the connected states. One has at this moment good hints of changing deformation of the ground state structure, but more data, preferentially leading to complementary information are needed for the Sr isotopes in extending the experimental knowledge beyond $N=100$ and moving further. Spectroscopy using fission at and near to nuclear reactors and separation of the fission products, which has been already carried out for the Zr isotopes are highly demanded for Sr isotopes too.  

\section{Acknowledgment}
We are very grateful to K.\ Heyde for the careful reading of this manuscript. This work was supported
by the grant number PID2019-104002GB-C21 funded by MCIN/AEI/10.13039/501100011033 and FEDER ``A way
of making Europe'', the Consejer\'{\i}a de Econom\'{\i}a, Innovaci\'on, Ciencia y Empleo de la Junta de Andaluc\'{\i}a (Spain) under Group FQM-370, and by FEDER  SOMM17/6105/UGR. Resources supporting this work were provided by the CEAFMC and the Universidad de Huelva High Performance Computer (HPC@UHU) funded by ERDF/MINECO project UNHU-15CE-2848.   

\bibliography{references-IBM-CM,references-QPT}
\end{document}